\documentclass[twocolumn,showpacs,preprintnumbers,amsmath,amssymb,unsortedaddress,nofootinbib]{revtex4}

\usepackage{amsfonts}
\usepackage{aas_macros}
\usepackage{graphicx}

\def\hmpc{\ {\rm {\it h}^{-1}Mpc}}

\def\hmmpc{\ {\rm {\it h}Mpc^{-1}}}
\def\hhmmpc{\ {\rm {\it h}^{2}Mpc^{-2}}}

\def\hhhgpc{\ {\rm {\it h}^{-3}Gpc^3}}
\def\mdh{\ {\rm M_\odot/{\it h}}}

\def\la{\langle}
\def\ra{\rangle}

\def\dc{\delta_{\rm c}}

\def\fnl{f_{\rm NL}}
\def\gnl{g_{\rm NL}}
\def\ek{\epsilon_\kappa}
\def\ei{\epsilon_{\rm I}}

\def\am{\alpha_M}

\def\vk{{\bf k}}
\def\vx{{\bf x}}
\def\vr{{\bf r}}

\def\nn{{\bf\hat{n}}}
\def\kk{{\bf\hat{k}}}

\def\mdh{\ {\rm M_\odot/{\it h}}}

\def\pmm{P_{\rm mm}}
\def\pmh{P_{\rm mh}}
\def\phh{P_{\rm hh}}
\def\nh{\bar{n}_{\rm h}}

\pacs{98.65.-r,~98.80.Cq,~95.36.+x,~98.70.Vc}

\begin{document}

\title{Signature of primordial non-Gaussianity of $\phi^3$-type 
in the mass function and bias of dark matter haloes}

\author
{Vincent Desjacques$^1$\thanks{dvince@physik.uzh.ch} and 
Uro\v s  Seljak$^{1,2,3}$\thanks{seljak@physik.uzh.ch} \\ 
$^1$ Institute for Theoretical Physics, University of Z\"urich, 
Winterthurerstrasse 190, CH-8057 Z\"urich, Switzerland \\ 
$^2$ Physics and Astronomy Department, University of California, and 
Lawrence Berkeley National Laboratory, \\ Berkeley, California 94720, USA \\
$^3$ IEU, Ewha University, Seoul, S. Korea}
\email[]{dvince@physik.uzh.ch}
\email[]{seljak@physik.uzh.ch}


\begin{abstract}

We explore the effect of a cubic correction $\gnl\phi^3$ on the mass
function and bias of dark matter haloes extracted from a series of
large N-body simulations and compare  it to theoretical
predictions. Such cubic terms  can be motivated in scenarios like the
curvaton model, in which a large cubic correction can be produced
while simultaneously keeping the quadratic $\fnl\phi^2$ correction small.
The deviation from the Gaussian halo mass  function is in reasonable
agreement with the theoretical predictions. The scale-dependent bias
correction $\Delta b_\kappa(k,\gnl)$  measured from the auto- and
cross-power spectrum of haloes, is similar to the correction in $\fnl$
models,  but the amplitude is lower than theoretical expectations.
Using the compilation of LSS data in Slosar et al. [JCAP, {\bf 08},
031 (2008)], we obtain  for the first time a limit  on $\gnl$ of
$-3.5\times 10^5 < \gnl < +8.2 \times 10^5$ (at 95\% CL).  This limit
will improve with the future LSS data by 1-2 orders of  magnitude,
which should test many of the scenarios of this type.

\end{abstract}

\maketitle

\setcounter{footnote}{0}

\section{Introduction}
\label{sec:intro}

In standard single field inflation, primordial curvature perturbations
are produced by the inflaton field as it slowly rolls  down its
potential (\cite{1981JETPL..33..532M,1982PhLB..117..175S,
1982PhLB..115..295H,1982PhRvL..49.1110G}). Most
of these models predict a nearly scale-invariant spectrum of adiabatic
curvature fluctuations in agreement with cosmological observations. In
addition, very small deviations from Gaussianity are expected
\cite{1987PhLB..197...66A,1992PhRvD..46.4232F,1994ApJ...430..447G}.
Therefore, any evidence for or against the detection of primordial
non-Gaussianity would strongly constrain inflationary scenarios.

Non-Gaussianity can be produced by nonlinearities in the relation
between the primordial curvature perturbation $\Phi$ (Here and
henceforth, the usual Bardeen potential in matter-dominated era) and
the inflaton field, interactions of scalar fields, a modified
dispersion relation or a  departure from the natural adiabatic vacuum
state (see \cite{2004PhR...402..103B} for a review).  Any
non-Gaussianity that is generated outside the horizon induces a
three-point function (or bispectrum) $B_\Phi(\vk_1,\vk_2,\vk_3)$ that
is peaked on squeezed triangles (i.e. $k_1\ll k_2\sim k_3$) for
realistic values of the scalar spectral index. The resulting
non-Gaussianity depends only on the local value $\Phi(\vx)$ of the
Bardeen's curvature potential and can thus be conveniently
parametrised up to third order by
\begin{equation}
\Phi(\vx)=\phi(\vx)+\fnl\bigl[\phi^2(\vx)-\la\phi^2\ra\bigr]+\gnl
\phi^3(\vx)\;,
\label{eq:phiexp}
\end{equation} 
where $\phi(\vx)$ is an isotropic Gaussian random field and $\fnl$,
$\gnl$ are dimensionless, phenomenological parameters. While the
quadratic term generates the irreducible three-point function or
bispectrum at leading order, the cubic term does so for the
irreducible four-point function or trispectrum. These statistics can
be computed straightforwardly from a perturbative expansion of the
homogeneous Robertson-Walker background
\cite{2003NuPhB.667..119A,2003JHEP...05..013M}. Convolved with  the
appropriate transfer function (e.g. the radiation transfer function
for the CMB temperature anisotropy), they can be  used to constrain
the value of the coupling parameters $\fnl$ and $\gnl$. No significant
detection of primordial non-Gaussianity has been reported from
measurements of the three-point correlation function of the cosmic
microwave background (CMB) anisotropies
\cite{2003ApJS..148..119K,2007JCAP...03..005C,2009ApJS..180..330K,
2009arXiv0901.2572S,2009MNRAS.393..615C}.  The tightest limits are
$-4<\fnl< 80$ at 95\% confidence level \cite{2009arXiv0901.2572S}.

If ${\cal O}(\fnl)\sim{\cal O}(\gnl)$ then  the cubic correction
should  always be negligibly small compared to the quadratic one since
curvature perturbations are typically ${\cal O}(10^{-5})$.  However,
this condition is not satisfied by some multifield inflationary models
such as the curvaton scenario, in which a large $\gnl$ and a small
$\fnl$ can be simultaneously produced. In this model, curvature
perturbations are generated by an additional scalar field, the
curvaton, whose energy density is negligible  during inflation
\cite{2003PhRvD..67b3503L,2004PhRvD..69d3503B,2005JCAP...10..013E,
2006JCAP...09..008M}.  Non-Gaussianity is generated by curvaton
self-interactions which effectively contribute a non-quadratic term to
the curvaton  potential
\cite{2006PhRvD..74j3003S,2008JCAP...09..012E,2008JCAP...09..025H,
2008PhRvD..78b3513I,2009JCAP...04..031C}.  While the value and  the
sign of $\gnl$ depend upon the exact form of the self-interaction term
(which can dominate the mass term if the curvaton mass is small enough
and the curvaton vacuum expectation value during inflation is large
enough \cite{2008JCAP...11..005H}), it is generically of magnitude
$|\gnl|\sim 10^4-10^5$ for realistic models in which the ratio of the
energy density of the curvaton to the total energy density at time of
decay is small. There are other realizations where one can have large
$\gnl$ and small $\fnl$
\cite{2009JCAP...06..035H,2009JCAP...08..016B}.  In ekpyrotic and
cyclic models, $\fnl$ typically is of the order of a few tens while
$\gnl$ is of the order of a few thousand \cite{2009PhRvD..80f3503L}.
If $\fnl$ were small, then the imprint of non-Gaussianity would be
detected only in four-point statistics such as the CMB trispectrum
\cite{2001PhRvD..64h3005H,2002PhRvD..66f3008O,2006PhRvD..73h3007K}. Thus
far, no observational limits have been set on $\gnl$ by measuring the
CMB trispectrum \cite{2002astro.ph..6039K,2001ApJ...563L..99K}.
Nevertheless, since the current bound $|\fnl|\lesssim 100$ implies a
relative contribution for the quadratic term of $\sim 0.1$ per cent, a
third order coupling parameter $|\gnl|\sim 10^6$ should also be
consistent with the data.

Large-scale structures offer another venue to test for the presence
of primordial non-Gaussianity. Deviation from Gaussianity can
significantly  affect the high mass end of the mass function
\cite{1988ApJ...330..535L,1989ApJ...345....3C}, the large-scale 
two-point correlation \cite{1986ApJ...310...19G,1986ApJ...310L..21M}, 
the bispectrum
\cite{2004PhRvD..69j3513S,2007PhRvD..76h3004S,2009arXiv0905.0717S,
2009arXiv0904.0497J}
of dark matter haloes hosting the observed galaxies as well as void
abundances \cite{2009JCAP...01..010K,2009MNRAS.395.1743L} and
topological measures of the cosmic web 
\cite{2006ApJ...653...11H,2008MNRAS.385.1613H}. Recently, references
\cite{2008PhRvD..77l3514D,2008ApJ...677L..77M,2008JCAP...08..031S}
showed that the local quadratic coupling $\fnl\phi^2$ induces a
scale-dependent bias $\Delta b_\kappa(k,\fnl)$ in the large-scale
power spectrum of biased tracers,
\begin{equation}
\Delta b_\kappa(k,\fnl)= 3\fnl \bigl[b(M)-1\bigr]\dc\frac{\Omega_{\rm
m}H_0^2}{k^2 T(k) D(z)}\;,
\label{eq:fnlshift}
\end{equation}
where $b(M)$ is the linear bias parameter, $H_0$ is the Hubble
parameter, $T(k)$ is the matter transfer function normalised to unity
as $k\rightarrow 0$, $D(z)$ is the growth  factor normalised to
$(1+z)^{-1}$ in the matter era and $\dc \sim 1.68$ is the present-day
(linear) critical density threshold.  Reference
\cite{2008JCAP...08..031S} applied Eq.~(\ref{eq:fnlshift}) to
constrain the value of $\fnl$ using a compilation of large-scale
structure data and found $-29<\fnl<+69$ at 95\% confidence. These
limits are comparable with those from the CMB, demonstrating the
competitiveness of the method.  Forthcoming all sky surveys should
achieve constraints of the order of $\fnl\sim 1$
\cite{2008PhRvD..77l3514D,2008PhRvD..78l3507A,2008ApJ...684L...1C,2009PhRvL.102b1302S} 
and should be sensitive to a possible scale-dependence of $\fnl$
\cite{2009arXiv0906.0232S}.
On the numerical side however, while simulations of structure
formation have confirmed the scaling $\Delta b_\kappa(k,\fnl)$ with
$k$
\cite{2008PhRvD..77l3514D,2009MNRAS.396...85D,2008arXiv0811.4176P,2009arXiv0902.2013G},
the exact amplitude of the non-Gaussian bias correction remains
somewhat debatable.

All numerical studies to date have only implemented the quadratic term
$\fnl\phi^2$.  The purpose of this work is to quantify the impact of
the cubic term $\gnl\phi^3$ on the mass function and bias of dark
matter haloes extracted from cosmological simulations and assess the
ability of  forthcoming measurements of the large-scale bias of
galaxies/quasars to constrain  the size of a local cubic
correction. This paper is organized as follows. We begin with a brief
description of the N-body simulations and illustrate the extent to
which the  coupling $\gnl\phi^3$ affects the matter power spectrum and
the halo mass function (Sec. \ref{sec:nbody}). We pursue with the
non-Gaussian halo bias (Sec. \ref{sec:biasgnl}), to which we derive
analytically the scale-dependent and scale-independent contribution,
$\Delta b_\kappa$ and $\Delta b_{\rm I}$, and demonstrates  the large
suppression of the simulated $\Delta b_\kappa$ relative to theory. We
then place limits  on the  coupling parameter $\gnl$ and forecast
constraints from future  large-scale surveys and CMB experiments
(Sec. \ref{sec:bound}). We also show that our  findings consistently
apply to more general models with non-zero $\fnl$ and $\gnl$
(Sec. \ref{sec:biasfnlgnl}). We conclude with a discussion of the
results in Sec. \ref{sec:discussion}.

\section{The non-Gaussian simulations}
\label{sec:nbody}

\subsection{Characteristics of the N-body runs}

We utilize a series  of large N-body simulations of the $\Lambda$CDM
cosmology seeded with  Gaussian and non-Gaussian initial
conditions. The (dimensionless) power spectrum of the Gaussian part
$\phi(\vx)$ of the Bardeen potential is the usual power-law
$\Delta_\phi^2(k)\equiv  k^3 P_\phi(k)/(2\pi^2)=A_\phi
(k/k_0)^{n_s-1}$.  The non-Gaussianity  is of the ``local'' form
$\Phi=\phi+\gnl\phi^3$.  We adopt the standard (CMB) convention in
which $\Phi(\vx)$ is primordial, and not extrapolated to present
epoch.  It is important to note that the local transformation is
performed before multiplication by the matter transfer
function. $T(k)$ is computed with {\small CMBFAST}
\cite{1996ApJ...469..437S}  for the WMAP5 best-fitting parameters
\cite{2009ApJS..180..330K}~:  $h=0.7$, $\Omega_{\rm m}=0.279$,
$\Omega_{\rm b}=0.0462$, $n_s=0.96$ and a normalisation of the
Gaussian curvature perturbations $A_\phi=7.96\times 10^{-10}$ at the
pivot point $k_0=0.02$Mpc$^{-1}$. This yields a density fluctuations
amplitude $\sigma_8\approx 0.81$ when the initial conditions are
Gaussian. Five sets of three 1024$^3$ simulations, each of which has
$\gnl=0,\pm 10^6$, were run with the N-body code {\small GADGET2}
\cite{2005MNRAS.364.1105S}.  We used the same Gaussian random seed
field $\phi$ in each set of runs so as to minimise the sampling
variance.  We also explored scenarios with non-zero $\fnl$ and $\gnl$
and ran 2  realisations   for each of the non-Gaussian models
characterized by $(\fnl,\gnl)=(\pm 100, -3\times 10^5)$. In all cases,
the box size is 1600$\hmpc$ with a force resolution of 0.04 times the
mean interparticle distance.  The particle mass of these simulations
thus is $3.0\times 10^{11}\mdh$, enough to resolve haloes down to
$10^{13}\mdh$.

In the curvaton scenario, generic polynomial interaction terms of the
form $\lambda m_\sigma^4 (\sigma/m_\sigma)^n$ (where $\lambda$ is some
coupling strength and $m_\sigma$ is the curvaton mass) to the
quadratic potential of the curvaton field $\sigma$ yield $|\gnl|\gg 1$
even when the non-linearity parameter $\fnl$ is very small
\cite{2008JCAP...09..012E,2008JCAP...09..025H}.  One   typically finds
$|\gnl|\sim {\cal O}(10^4) - {\cal O}(10^5)$ when $\fnl$ varies in the
range $-100<\fnl<100$. For practical  reasons however, the values of
$\gnl$ adopted in our simulations are about an order of magnitude
larger so as to produce an effect strong enough to be unambiguously
measured despite the small simulated volume. Furthermore, we have also
considered positive and negative values of $\gnl$ so as to assess the
sensitivity of  the non-Gaussian bias to the sign of the coupling
parameter.  The simulations with $(\fnl,\gnl)=(-100,-3\times 10^5)$
may be seen as a particular realisation of the curvaton model in which
the coupling constant $\lambda$ is positive, and the  non-quadratic
term is very steep ($n\sim 5-10$) but contributes little to the total
curvaton potential.

\subsection{Properties of the initial density field}

In order to ensure that the initial conditions are successfully
generated, we measure at the redshift of our initial conditions,
$z=99$,  the (normalized) skewness
$S_3(R,z)=\la\delta_{R,z}^3\ra/\sigma^4$ and kurtosis
$S_4(R,z)=(\la\delta_{R,z}^4\ra/-3\sigma^4)/\sigma^6$ of the density
field $\delta_{R,z}$ smoothed with a (spherically symmetric) window
function of characteristic radius $R$.  We adopt a tophat filter
throughout this paper. Note also that $\sigma(R,z)$ is the variance of
smoothed density fluctuations at redshift $z$.

In the weakly nonlinear regime, the skewness and kurtosis of the
density field may be written as the sum of a part generated by
gravitational clustering and a part  induced by primordial
non-Gaussianity. For Zel'dovich initial  conditions
\cite{1970A&A.....5...84Z} and $\Omega_m(z)\approx 1$, the
contribution generated by gravitational instabilities reads as
\cite{1984ApJ...279..499F,1986ApJ...311....6G,1992ApJ...394L...5B,
1994ApJ...433....1B}
\begin{align}
  S_3^{\rm Zel}(R,z) &= 4-\left(n_{\rm eff}+3\right) \label{eq:za} \\  
  S_4^{\rm Zel}(R,z) &= \frac{272}{9}-\frac{50}{3}\left(n_{\rm eff}+3\right)
  +\frac{7}{3}\left(n_{\rm eff}+3\right)^2 \nonumber\;,
\end{align}
where $n_{\rm eff}(R)$ is the effective spectral index at the
smoothing  scale $R$,
\begin{equation}
n_{\rm eff}(R) \equiv -\frac{d\ln\sigma^2(R,z)}{d\ln R}-3\;.
\end{equation}
Note that the initial skewness and kurtosis given by the Zel'dovich
approximation differs from the exact values predicted  by perturbation
theory, to which they asymptote in the limit $D(z)\to\infty$
\cite{1998MNRAS.299.1097S}.  In addition, the cubic coupling
$\gnl\phi^3$ induces a nonzero kurtosis $S_4^{\rm Pri}(R,z)\equiv\gnl
S_4^{(1)}(R,z)$ at leading   order which can be computed analytically
from the relation
\begin{align}
\sigma^6 S_4^{(1)}(R,z) &= 4!\left( \prod_{i=1}^{3} \int\!\!\frac{d^3
k_i}{(2\pi)^3}\,\alpha_R(k_i,z) P_\phi(k_i)\right)  \nonumber \\ &
\quad \times \alpha_R\bigl(|\vk_1+ \cdots + \vk_{n-1}|,z\bigr)\;,
\label{eq:moments}
\end{align}
where
\begin{equation}
\alpha_R(k,z)=\frac{2}{3\Omega_{\rm m}H_0^2}D(z) k^2 T(k) W_R(k)
\label{eq:alpha}
\end{equation} 
is evaluated at redshift $z$ and $W_R(k)$ is the Fourier transform of
the  tophat function. When $\fnl=0$, primordial skewness is not
generated at the first order and, therefore, may be neglected.

\begin{figure}
\center \resizebox{0.5\textwidth}{!}{\includegraphics{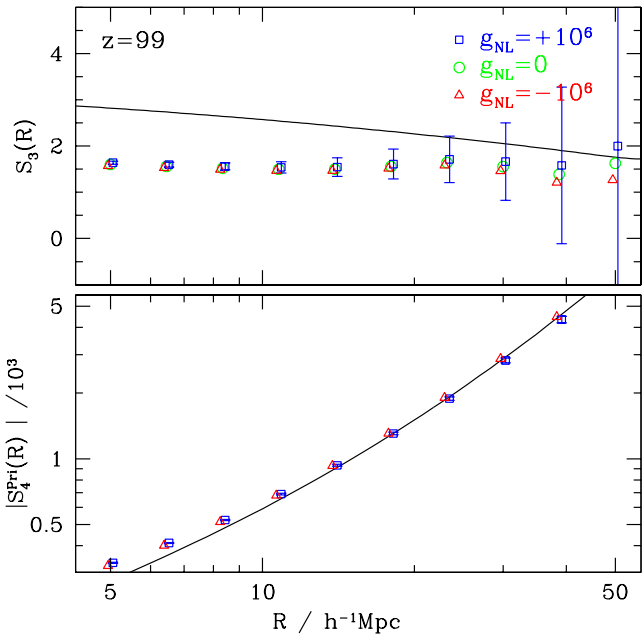}}
\caption{Skewness and kurtosis of the initial ($z=99$) density field
as a function of smoothing radius. While the top panel shows
the sum of the contributions arising from the Zel'dovich dynamics and
from primordial non-Gaussianity, $S_3=S_3^{\rm Zel}+S_3^{\rm Pri}$,
the bottom panel only shows the absolute value of the primordial
kurtosis, $|S_4^{\rm Pri}|$. Symbols represent the numerical results
averaged over the realisations. They have been slightly shifted 
horizontally for clarity. Error bars show the scatter among the
realisations for the models with $\gnl=10^6$. 
Solid lines indicate the theoretical expectations.}
\label{fig:ics}
\end{figure}

Fig.~\ref{fig:ics} displays the initial skewness (top panel) and
kurtosis (bottom panel) obtained upon distributing the dark matter
particles onto a regular 512$^3$ mesh (i.e. of cell size $\approx
3\hmpc$) using the cloud-in-cell (CIC)  interpolation scheme. Symbols
represent the numerical results averaged  over the realisations.
Because $S_4^{\rm Zel}$ (which is the sole contribution to the
kurtosis in the Gaussian case) varies considerably between the
realisations,  we only show the absolute difference $|S_4^{\rm Pri}|$
between the kurtosis in the non-Gaussian ($\gnl=\pm 10^6$)  and the
Gaussian ($\gnl=0$) runs. As expected, the skewness is very similar
among the  Gaussian and non-Gaussian simulations. While $S_3$ in the
simulations  agrees well with the skewness induced by the Zel'dovich
dynamics Eq.  (\ref{eq:za}) (solid curve), it is gradually suppressed
as the filtering radius $R$ approaches the cell size, presumably
because of the finite resolution which smoothes also the fluctuations.
Note that, in the simulations with  nonzero $\gnl$ and $\fnl$, there
is a large primordial skewness which is of magnitude $|S_3^{\rm
Pri}|\sim$ a few on scale  $R\lesssim 50\hmpc$, consistent with
theory.  As can also be seen, the absolute value of the primordial
kurtosis increases sharply with $R$, in very good agreement with the
theoretical prediction.

\subsection{The matter power spectrum}
    
Non-Gaussian corrections to the primordial curvature perturbation can
renormalise the input power spectrum of fluctuations used to seed the
simulations. Since our simulations implement its unrenormalised
version $\Delta_\phi^2(k)=A_\phi (k/k_0)^{n_s-1}$, it is desirable to
ascertain the effect of the local coupling term on the simulated
density power spectrum before discussing the halo mass function and
bias. For $\fnl$ models with $|\fnl|\lesssim 100$, renormalisation
effects are unlikely to be noticeable due to the limited dynamical
range of current cosmological simulations
\cite{2008PhRvD..78l3519M}. As we will see shortly however, they can
be significant in simulations of $\gnl$ models with similar level of
non-Gaussianity.

The cubic order term $\gnl\phi^3$ renormalises the amplitude $A_\phi$
of the power spectrum of initial curvature perturbations to
$A_\phi\rightarrow A_\phi+6\gnl\la\phi^2\ra$, where
\begin{equation}
\la\phi^2\ra=\int\!\!\frac{d^3k}{(2\pi)^3}P_\phi(k)\;.
\end{equation}
For scale invariant initial conditions, $\la\phi^2\ra$ has a
logarithmic divergence at large and small scales (see
\cite{2008PhRvD..78l3519M} for a more detailed discussion of
perturbative corrections in non-Gaussian cosmologies).  In practice, a
low- and high-$k$ cutoff are naturally provided by the finite box size
and the resolution of the simulations.  Therefore, the effective
amplitude of density fluctuations in non-Gaussian simulations with
cubic coupling  is $\sigma_8+\delta\sigma_8$ with
\begin{align}
\delta\sigma_8 &= 3\gnl\la\phi^2\ra \\
 &= 3\gnl \left(\frac{k_0}{k_{\rm min}}\right)^{1-n_s}
\left[1-\left(\frac{k_{\rm min}}{k_{\rm max}}\right)^{1-n_s}\right]
\frac{A_\phi}{1-n_s} \nonumber\;.
\end{align}
Recall that $k_0=0.02\hmmpc$ is our choice of normalisation point, and
$k_{\rm min}$ and $k_{\rm max}$ are the integration limits set by  the
fundamental mode and the Nyquist frequency of the periodic cubical
box over which the initial conditions are generated. Equivalently,
\begin{equation}
\delta\sigma_8=3\gnl
\left(\frac{L k_0}{2\pi}\right)^{1-n_s}\left[1-N^{n_s-1}\right]
\frac{A_\phi}{1-n_s}\;,
\end{equation}
where $N=1024$ is the number of mesh points along one dimension. This
result becomes $\delta\sigma_8=3\gnl\ln(N) A_\phi$
in the scale-invariant limit $n_s\rightarrow 1$. For
the cosmological setup considered here, the absolute deviation is
\begin{equation}
\delta\sigma_8\approx 0.015\left(\frac{\gnl}{10^6}\right)\;.
\label{eq:dsigma8}
\end{equation} 
This correction is fairly large  for the values of $\gnl$ adopted here
and, therefore, must be taken into account  in the comparison between
the theory and the simulations. As we will see below, this is
especially important when studying the high mass tail of the halo mass
function which is exponentially sensitive to the amplitude of density
fluctuations.

\begin{figure}
\center \resizebox{0.5\textwidth}{!}{\includegraphics{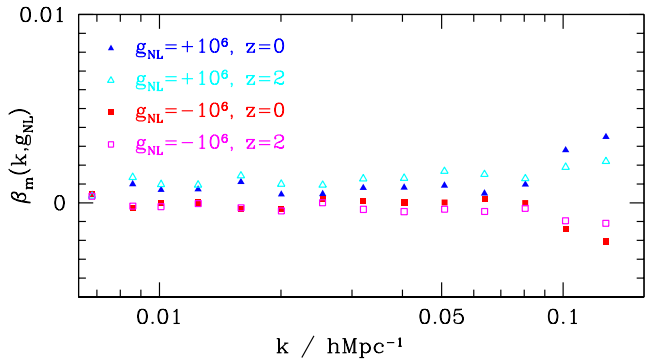}}
\caption{Non-Gaussian fractional correction $\beta_{\rm
m}(k,\gnl)=P_{\rm mm}(k,\gnl)/P_{\rm mm}(k,0)-1$ to the matter power
spectrum after subtracting a scale-independent normalisation shift
$6\gnl\la\phi^2\ra$ induced by the cubic coupling $\gnl\phi^3$.}
\label{fig:matter}
\end{figure}

The cubic coupling term $\gnl\phi^3$ can also induce a scale-dependent
correction to the matter power spectrum which can be quantified by the
fractional change $\beta_{\rm
m}(k,\gnl)=\pmm(k,\gnl)/\pmm(k,\gnl=0)-1$.  In Fig.~\ref{fig:matter},
symbols show the result of measuring  $\beta_{\rm m}(k,\gnl)$ from the
snapshots at $z=0$ and 2 after correction of the normalisation shift
$|2\delta\sigma_8/\sigma_8|=0.037$.  There is some evidence for a
scale-dependent correction at wavenumber $k\gtrsim 0.1\hmmpc$ but  the
resulting deviation is broadly consistent with zero. We will thus
neglect $\beta_{\rm m}(k,\gnl)$ henceforth.

\subsection{The halo multiplicity function}

Haloes were identified using the {\small MPI} parallelised version of
the {\small AHF} halo finder \cite{2009arXiv0904.3662K} which is based
on the spherical overdensity (SO) finder developed by
\cite{2004MNRAS.351..399G}. The main reason for using a SO finder is
that it is more closely connected to the predictions of the spherical
collapse model, on which most of the analytic formulae presented in
this paper are based. Namely, the virial mass  $M$ of a halo is
defined by the radius at which the inner overdensity exceeds
$\Delta_{\rm vir}(z)$ times the background density $\bar{\rho}(z)$
\cite{1992ApJ...399..405W,1994MNRAS.271..676L}.  The value of the
overdensity threshold $\Delta_{\rm vir}(z)$ is obtained from the
collapse of a spherical tophat perturbation and has a dependence on
redshift through the matter density $\Omega_m(z)$
\cite{1996MNRAS.282..263E,1998ApJ...495...80B}.  We discard poorly
resolved haloes and only study those  containing at least 34 particles
or, equivalently, with a mass larger than $M=10^{13}\mdh$.

Analytic arguments based on the Press-Schechter theory
\cite{1974ApJ...187..425P,1991ApJ...379..440B} predict that the halo
mass function $n(M,z)$ is entirely specified by the distribution  $\nu
f(\nu)$ of first-crossings, or multiplicity function
\begin{equation} 
\nu f(\nu)=M^2\,\frac{n(M,z)}{\bar{\rho}}\frac{d\ln M}{d\ln\nu}\;.
\label{eq:fnu}
\end{equation} 
The peak height $\nu(M,z)=\dc(z)/\sigma(M)$ is the typical amplitude
of fluctuations that produce haloes of mass $M$ by redshift $z$. Here
and henceforth, $\sigma(M)$ denotes the variance of the density field
$\delta_M$ smoothed on mass scale $M\propto R^3$ and linearly
extrapolated to present epoch, whereas $\dc(z)\approx 1.68 D(0)/D(z)$
is the critical linear overdensity for (spherical) collapse at
redshift $z$.

Despite the lack of a compelling theoretical description of the
multiplicity function for Gaussian initial conditions, the fractional
deviation from Gaussianity can be  modelled accurately using the
Press-Schechter formalism. In this approach, the halo mass function
$n(M,z)$ is related to the probability $P(>\dc,M)$ that a region of
mass $M$ exceeds the critical density for collapse $\dc(z)$ through
the  relation $n(M,z)=-2\,(\bar{\rho}/M)\,dP/dM$. The non-Gaussian
fractional correction to the multiplicity function then is
$R(\nu,\gnl)\equiv
f(\nu,\gnl)/f(\nu,0)=(dP/dM)(>\dc,M,\gnl)/(dP/dM)(>\dc,M,0)$. The
level excursion probability $P(>\dc,M,\gnl)$ can be computed once the
probability distribution function (PDF) of the smoothed density  field
$\delta_M$, $P(\delta_M)$, is known. Here, we will consider the simple
extensions proposed by \cite{2008JCAP...04..014L} and
\cite{2000ApJ...541...10M}, in which $P(\delta_M)$ is generically
expressed as the inverse transform of a cumulant generating function.
Both extensions have been shown to give reasonable agreement with
numerical simulations of non-Gaussian cosmologies
\cite{2007MNRAS.382.1261G,2009MNRAS.396...85D,2009arXiv0902.2013G}.

In \cite{2008JCAP...04..014L}, the saddle-point technique is applied
directly to $P(\delta_M)$. The resulting Edgeworth expansion is then
used to obtain $P(>\dc,M,\gnl)$. For $\fnl$ non-Gaussianity, reference
\cite{2009MNRAS.396...85D} found that the resulting non-Gaussian mass
function agrees  well with the simulations.  For $\gnl$
non-Gaussianity, neglecting cumulants other than the kurtosis
$S_4^{\rm Pri}(M)$ (Hereafter, we drop the superscript for
conciseness) and truncating the series expansion at $S_4(M)$, the
non-Gaussian fractional correction reads
\begin{align}
 R_{\rm LV}(\nu,\gnl) &\approx  
 \left\{1+\frac{1}{4!}\,\sigma^2 S_4\left(\nu^4-4\nu^2-3\right)\right. 
 \\  & \qquad \left.-\frac{1}{4!}\sigma^2\frac{d S_4}{d\ln\nu}
 \left(\nu^2-3\right)
 \right\}\exp\left[\nu^2\delta\sigma_8\right] \nonumber \\ &=
 \left\{1+\frac{1}{4!}\,\sigma^2 S_4 \left(\nu^4-6\nu^2+3\right)\right.
 \label{eq:loverde}
 \\  & \qquad \left. -\frac{1}{4!}\, \frac{d(\sigma^2S_4)}{d\ln\nu}
 \left(\nu^2-3\right)\right\}\exp\left[\nu^2\delta\sigma_8\right]
\nonumber
\end{align} 
after integration over regions above  $\dc(z)$. Note that we have
omitted writing the redshift dependence explicitly.  Strictly speaking
however, $R(\nu,\gnl)$ depends distinctly upon the variables $M$ (or
$\nu$) {\it and} $z$ due to the presence of $\sigma^2 S_4(M)$. Our
notation is motivated by the fact that the measured non-Gaussian
correction, as plotted in Fig.\ref{fig:fnu}, appears to depend mostly
on the peak height. The exponential factor in the right-hand side is
the correction induced by the renormalisation of the amplitude of
linear density fluctuations, Eq.~(\ref{eq:dsigma8}). For consistency,
we have also used the Press-Schechter multiplicity function to derive
this last term although  a Sheth-Tormen mass function
\cite{1999MNRAS.308..119S} may be more appropriate.

In \cite{2000ApJ...541...10M}, it is the level excursion probability
$P(>\dc,M)$ that is calculated within the saddle-point approximation.
Including only a cubic coupling $\gnl\phi^3$ and truncating the
resulting expression at the kurtosis, we  find
\begin{align}
\lefteqn{P(>\dc,M,\gnl)} \\
 &\quad\approx\frac{1}{\sqrt{2\pi}}\frac{\sigma}{\dc}
  \left(1+3\gnl\sigma^2\la\phi^2\ra-\frac{S_4}{12}\sigma^2\dc^2\right) 
  \nonumber \\
 & \qquad \times \exp\left\{-\frac{\dc^2}{2\sigma^2}\left[1-6\gnl
    \la\phi^2\ra-\frac{S_4}{12}\,\dc^2\right]\right\}\nonumber
\end{align}
at first order in $\gnl$.  Note that we have already included the
renormalisation of the  fluctuation amplitude. For rare events,
$\sigma\ll 1$ and the first parenthesis in the right-hand side can be
neglected. To ensure that the resulting mass function is properly
normalised, we follow \cite{2009arXiv0906.1042V} and use
\begin{equation}
\nu_\star f(\nu_\star)=M^2\,\frac{n(M,z,\gnl)}{\bar{\rho}}\frac{d\ln M}
{d\ln\nu_\star}\;.
\end{equation}
for the non-Gaussian mass function, where $\nu_\star=\delta_\star/\sigma$,
$\delta_\star=\dc\sqrt{1-2\delta\sigma_8-S_4\dc^2/12}$~\footnote{
In local $\fnl$ models, this formula only involves the skewness $S_3$.
As it is incorrectly quoted in some of the literature on non-Gaussian
halo mass functions, let us write down its explicit expression:
\begin{equation}
R_{\rm MVJ}(\nu,\fnl)=\exp\left[\frac{S_3\dc^3}{6\sigma^2}\right]
\left[\frac{\dc^2}{6\delta_\star}\frac{d S_3}{d\ln\sigma}
+\frac{\delta_\star}{\dc}\right] \nonumber\;,
\end{equation} 
or, in terms of the peak height $\nu=\dc/\sigma$, 
\begin{equation}
R_{\rm MVJ}(\nu,\fnl)\approx\exp\left[\frac{\nu^3}{6}\sigma S_3\right]
\left[1-\frac{\nu}{3}\sigma S_3-\frac{\nu}{6}\frac{d(\sigma S_3)}
{d\ln\nu}\right] \nonumber\;,
\end{equation}
after expanding $\delta_\star=\sqrt{1-S_3\dc /3}$ at the first order.}
and $f$ is the same multiplicity function as in the Gaussian case. 
Taking the derivative of the level excursion probability then gives
\begin{align}
\frac{(dP/dM)(>\dc,M,\gnl)}{(dP/dM)(>\dc,M,0)} &\approx 
\exp\left[\frac{S_4\dc^4}{4!\sigma^2}+\nu^2\delta\sigma_8\right] \\ 
  & \quad \times \left(\frac{\delta_\star}{\dc}   
     +\frac{1}{4!}\frac{\dc^3}{\delta_\star}\frac{dS_4}{d\ln\sigma}
     \right)\nonumber \;.
\end{align}
The fractional change in the multiplicity function eventually reads as
\begin{align}
\label{eq:mvj}
R_{\rm MVJ}(\nu,\gnl) &\approx
\exp\left[\frac{\nu^4}{4!}\sigma^2 S_4+\nu^2\delta\sigma_8\right] \\
 & \quad\times\left\{1-\frac{\nu^2}{8}\sigma^2 S_4-\frac{\nu^2}{4!}
    \frac{d(\sigma^2 S_4)}{d\ln\nu}\right\}\nonumber
\end{align}
after expanding $\delta_\star$ at the first order and ignoring the 
shift in the normalisation amplitude, i.e.  $\delta_\star^{\pm
1}\approx 1 \mp S_4\dc^2/24$.  In the limit $\sigma^2S_4\ll 1$ and
$\nu\gg 1$, the two theoretical expectations reduce to
$1+\nu^4\sigma^2S_4/24$. However, they differ in the coefficient of
the $\nu^2\sigma^2 S_4$ term, which is $-1/4$ and $-1/8$ for the LV
and MVJ formula, respectively. Therefore, we shall also consider the
approximation
\begin{align}
\label{eq:thiswork}
R(\nu,\gnl) &=
\exp\left[\frac{\nu^4}{4!}\sigma^2 S_4+\nu^2\delta\sigma_8\right] \\
 & \quad\times\left\{1-\frac{\nu^2}{4}\sigma^2 S_4-\frac{\nu^2}{4!}
    \frac{d(\sigma^2 S_4)}{d\ln\nu}\right\}\nonumber\;,
\end{align}
which is designed to match better the Edgeworth expansion of
\cite{2008JCAP...04..014L} when the peak height is $\nu\sim 1$. 

\begin{figure}
\center \resizebox{0.5\textwidth}{!}{\includegraphics{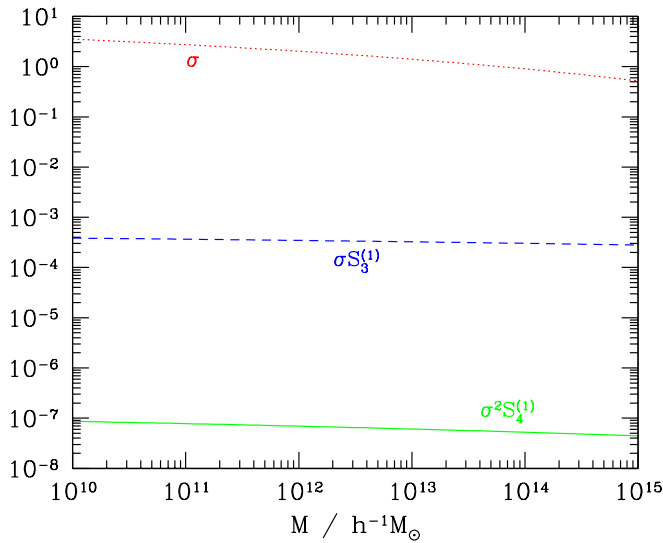}}
\caption{Variance $\sigma$ (dotted), skewness $\sigma S_3^{(1)}$ 
(dashed) and kurtosis $\sigma^2 S_4^{(1)}$ (solid) of the smoothed 
linear density field $\delta_M$ as a function of mass scale $M$.}
\label{fig:moments}
\end{figure}

\begin{figure}
\center \resizebox{0.5\textwidth}{!}{\includegraphics{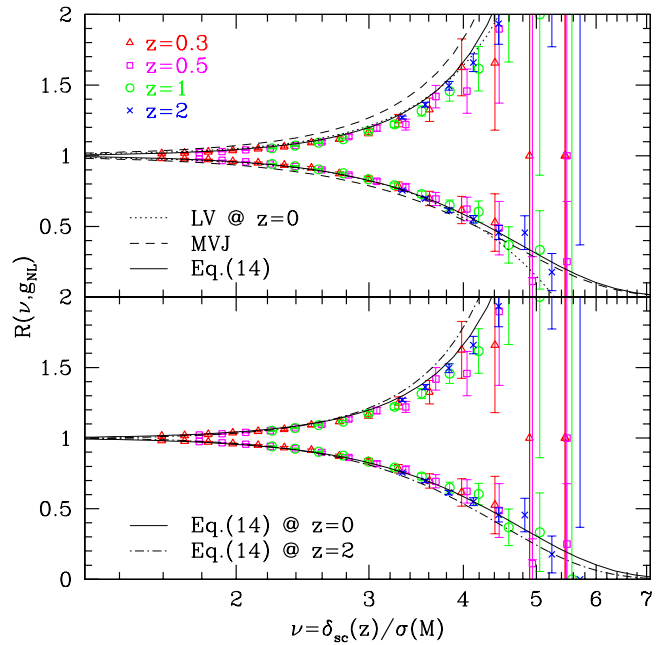}}
\caption{{\it Top panel}~: Fractional correction to the Gaussian
multiplicity function of dark matter haloes as a function of the peak
height $\nu(M,z)$ for a coupling parameter $\gnl=\pm 10^6$. The
dotted, dashed and solid curves show the theoretical predictions
eqs. (\ref{eq:loverde}), (\ref{eq:mvj}) and (\ref{eq:thiswork}) at
$z=0$, respectively. Error bars denote Poisson errors. For
illustration, $M=10^{15}\mdh$ corresponds to $\nu=3.2$, 5.2, 7.7 at
redshift $z=0$, 1 and 2, respectively. Similarly, $M=10^{14}\mdh$ and
$10^{13}\mdh$ correspond to $\nu=1.9$, 3, 4.5 and 1.2, 1.9, 2.9
respectively.  {\it Bottom panel}~: a comparison with equation
(\ref{eq:thiswork}) evaluated at $z=0$ and 2.}
\label{fig:fnu}
\end{figure}

Calculating the fractional change in the mass function requires
knowledge of the kurtosis $S_4(M)\equiv\gnl S_4^{(1)}(M)$ of the
smoothed density field $\delta_M$, which we compute analytically 
using the general formula (valid for $n\geq 3$)
\begin{align}
\sigma^{2n-2}S_n^{(1)}(M) &= n!\left( \prod_{i=1}^{n-1}
\int\!\!\frac{d^3 k_i}{(2\pi)^3}\,\alpha_M(k_i) P_\phi(k_i)\right) 
 \nonumber \\
 & \quad \times \alpha_M\bigl(|\vk_1+ \cdots + \vk_{n-1}|\bigr)\;,
\label{eq:moments}
\end{align}
where $\am(k)\equiv \alpha_R(k,z=0)$ in what follows.  Over the mass
range probed by our simulations, $10^{13}\lesssim M\lesssim 5\times
10^{15}\mdh$, the normalised kurtosis $\sigma^2 S_4^{(1)}(M)$ is a
monotonic decreasing function of $M$ that varies in the narrow range
$\sim 4-6\times 10^{-7}$ for the top-hat filter assumed here (see
Fig.~\ref{fig:moments}). Note also that the $\sigma^2 S_4$ term
dominates the total  contribution to the non-Gaussian correction
eqs. (\ref{eq:loverde}), (\ref{eq:mvj}) and (\ref{eq:thiswork}) when
the peak height is $\nu\gtrsim 2$ (One finds $|d(\sigma^2
S_4)/d\ln\nu|\lesssim 0.1 \sigma^2 S_4$).

The fractional correction is plotted in Fig. \ref{fig:fnu} for the
haloes extracted from the simulations at redshift $z=0.3$, 0.5, 1 and
2.  In the top panel, the data are compared to the theoretical
predictions eqs. (\ref{eq:loverde}),  (\ref{eq:mvj}) and
(\ref{eq:thiswork}) evaluated at $z=0$.  As we can see, the level of
non-Gaussianity in the halo multiplicity function is consistent with
the theory. Our approximation (\ref{eq:thiswork}) performs equally
well regardless of the sign of $\gnl$. It agrees better with the
measurements than the formulae of \cite{2000ApJ...541...10M} which
significantly overestimates the data for $\gnl=10^6$, and than that of
\cite{2008JCAP...04..014L} which is not always positive definite for
$\gnl=-10^6$. The bottom panel shows that the discrepancy somewhat
worsens at higher redshift, especially in the case
$\gnl=10^6$. However,  it is possible the agreement may be improved by
adding higher order powers of $\sigma^2 S_4$ and higher order
cumulants.

To conclude this section, one should keep in mind that all these
extensions are based on Press-Schechter and, therefore, provide a bad
fit to the Gaussian mass function of haloes. In this respect,
excursion set approaches may be more promising since they seem to
reproduce both the Gaussian halo counts and the dependence on $\fnl$
\cite{2009arXiv0905.1702L,2009MNRAS.399.1482L}.

\section{The non-Gaussian bias shift}
\label{sec:biasgnl}

\subsection{Theoretical considerations}
\label{sub:biastheory}

In order to calculate the scale-dependent bias correction induced by
the $\gnl$ coupling term to the correlation of haloes of mass $M$
collapsing at redshift $z$, we follow \cite{2008ApJ...677L..77M} and
consider the two-point  correlation $\xi_{\rm hh}(\vr)$ of regions of
the smoothed density field $\delta_M$ above a threshold
$\delta_c(z)=\nu(z)\sigma$. The two-point correlation function of
this level excursion set, which was originally derived by
\cite{1986ApJ...310L..21M}, can be expressed in the high threshold 
approximation as
\begin{align}
  \xi_{\rm hh}(\vr) &= -1 +
  \exp\left\{\sum_{n=2}^\infty\sum_{j=1}^{n-1}\frac{\nu^n\sigma^{-n}}
  {j!(n-j)!}\right. \\ & \qquad \left. \times
  \xi^{(n)}\!\left(\begin{array}{cc}\vx_1,\cdots,\vx_1,  &
  \vx_2,\cdots,\vx_2 \\ j~\mbox{times} & (n-j)~\mbox{times}\end{array}
  \right) \right\}  \nonumber \;,
\label{eq:corrlevel}
\end{align}
where $\vr=\vx_1-\vx_2$.  For the non-Gaussian model considered here,
the leading-order correction induced by non-zero three-point and
four-point correlations of the density field  reads
\begin{align}
\Delta\xi_{\rm hh} &=
\frac{\nu^3}{\sigma^3}\xi^{(3)}(\vx_1,\vx_1,\vx_2)+\frac{\nu^4}
{\sigma^4}\left[\frac{1}{3}\xi^{(4)}(\vx_1,\vx_1,\vx_1,\vx_2)\right.
\nonumber \\ & \qquad \left . 
+\frac{1}{4}\xi^{(4)}(\vx_1,\vx_1,\vx_2,\vx_2)\right]\;.
\end{align}
The non-Gaussian correction $\Delta P_{\rm hh}$ to the power spectrum
of biased tracers is obtained by Fourier transforming this expression.

In the case $\fnl=0$ and $\gnl\neq 0$, only the four-point functions
contribute at first order. It should also be noted that, at linear
order,  $\xi^{(2)}(\vx_1,\vx_2)$ amounts to a renormalisation of the
linear bias and, therefore, does not contribute to the scale-dependent
correction. Details of the calculation can be found in Appendix
\ref{app:ngbias}. In short, the non-Gaussian correction $\Delta P_{\rm
hh}$ in the limit of long wavelength $k\ll 1$  is given by the Fourier
transform of $\nu^4\xi^{(4)}(\vx_1,\vx_1,\vx_2,\vx_2)/3\sigma^4$,
\begin{align}
\label{eq:dphhg}
\Delta P_{\rm hh}(k) &= \gnl\nu^4\,S_3^{(1)}\!(M)\am(k) 
P_\phi(k) \\
&= \gnl b_{\rm L}^2(z) \delta_c^2(z)\, S_3^{(1)}\!(M)
\am(k) P_\phi(k) \nonumber \;,
\end{align}
where we have used $b_{\rm L}(z)=\nu^2/\delta_c(z)$ as is appropriate
for high density peaks. The smoothing window that appears in $\am(k)$
effectively makes little difference because we are considering the
limit where $k^{-1}$ is much larger than the smoothing radius, so we
will omit it in the following.  For small non-Gaussianity, we can also
write $\Delta P_{\rm hh}\approx 2 b_{\rm L}\Delta b_\kappa
P_\delta(k)$ where $P_\delta(k)=\am^2 P_\phi(k)$ is the power spectrum
of the smoothed density field. The scale-dependent bias correction
$\Delta b_\kappa(k,\gnl)$ can eventually  be recast into the  form
\begin{align}
\label{eq:gshiftk}
\Delta b_\kappa(k,\gnl) &= \frac{1}{2} \gnl b_{\rm L}(z)
\delta_c^2(z)\,S_3^{(1)}\!(M) \am^{-1}(k) \\
&= \frac{3}{4}\gnl b_{\rm L}(z)\delta_c^2(0)\frac{D(0)}{D(z)^2}
\,S_3^{(1)}\!(M)\frac{\Omega_{\rm m}H_0^2}{k^2 T(k)} \nonumber \\
&= \frac{1}{4}\gnl\delta_c(z)\,S_3^{(1)}\!(M)\,
\Delta b_\kappa(k,\fnl=1)\nonumber\;,
\end{align}
where $\Delta b_\kappa(k,\fnl)$ is the scale-independent bias induced
by the quadratic coupling $\fnl\phi^2$, Eq.~(\ref{eq:fnlshift}).  We
have also assumed the Eulerian bias prescription  $b(M)=1+b_{\rm
L}(M)$.

\begin{figure}
\center \resizebox{0.5\textwidth}{!}{\includegraphics{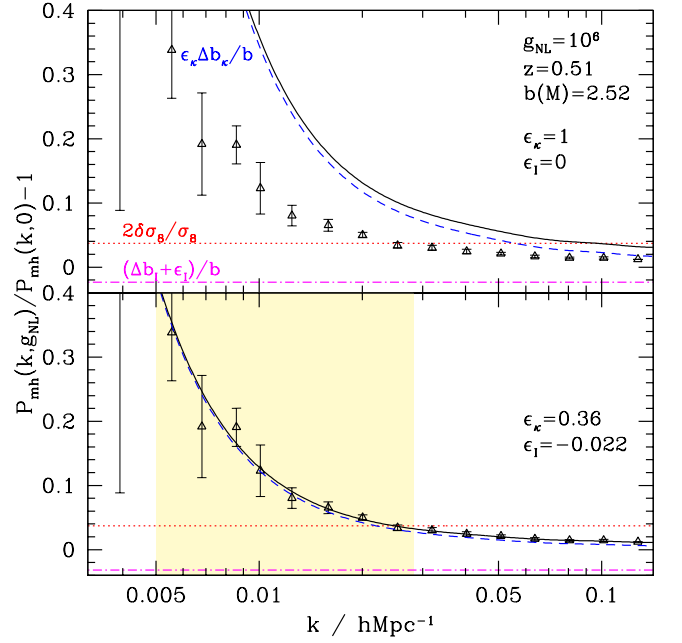}}
\caption{{\it Top panel}~: Non-Gaussian bias correction computed from
the halo-matter power spectrum of haloes of mass $M>2\times 10^{13}\mdh$
extracted from the snapshot at $z=0.5$ (filled symbols). The solid
curve represents $\pmh(k,\gnl)/\pmh(k,0)-1$ with a non-Gaussian bias
shift $\Delta b(k,\gnl)$ given by Eq.~({\ref{eq:dbias}}). The dashed,
dotted and dotted-dashed curves show three separate contributions that
arise at first order in $\gnl$. {\it Bottom panel}~: $\Delta b(k,\gnl)$
is replaced by the theoretical model Eq.~(\ref{eq:dbiaseff}).  The
shaded region indicates the  data points used to fit the parameters
$\ek$ and $\ei$. Error bars indicate the scatter among 5 realisations.}
\label{fig:dbias}
\end{figure}

The change in the mean number density of haloes also creates a
scale-independent shift which we denote by  $\Delta b_{\rm
I}(\gnl)$. As shown in \cite{2009MNRAS.396...85D} for $\fnl$ models,
the inclusion of this correction noticeably improves the agreement
with the simulations at wavenumber $k\lesssim 0.1\hmmpc$.  Using a
peak-background split and considering the limit of small
non-Gaussianity, this  contribution reads
\begin{align}
\Delta b_{\rm I}(\gnl) &=
-\frac{1}{\sigma}\frac{\partial}{\partial\nu}\ln\bigl[R(\nu,\gnl)
\bigr] \nonumber \label{eq:gshifti} \\ &\approx 
-\frac{1}{\sigma}\left[\frac{1}{3!}  \left(\nu^3-3\nu\right)\sigma^2
  S_4 + 2\nu\delta\sigma_8 \right. \\ & \qquad
  \left. +\frac{1}{4!}\left(\nu^3-8\nu\right)\frac{d(\sigma^2 S_4)}
       {d\ln\nu}-\frac{\nu}{4!}\frac{d^2(\sigma^2
	 S_4)}{d\ln\nu^2}\right] \nonumber
\end{align}
after truncating $\Delta b_{\rm I}$ at first order in $\gnl$ This
approximation should perform reasonably well for moderate values of
the peak height, $\nu\lesssim 4$, for which the fractional change in
the mass function, equation (\ref{eq:thiswork}), matches well the
numerical data. It is worth noticing that $\Delta b_{\rm I}(\gnl)$ has
a sign opposite to that of $\gnl$ because the bias decreases when the
mass function goes up. $\Delta b_{\rm I}(\gnl)$ also includes  a
correction induced by the renormalisation of $\sigma_8$.  In practice,
we estimate $\Delta b_{\rm I}(\gnl)$ for a given halo sample by
evaluating $\sigma^2 S_4$ and $\nu$ at the scale corresponding to the
average halo mass $\bar{M}$ of the sample, as it is unclear to which
extent $R(\nu,\gnl)$ agrees with the data in the limit $M\gg M_\star$.

\subsection{Comparison with the simulations}

To assess the effect of primordial  non-Gaussianity on the halo bias,
we will consider the ratios
\begin{align}
\frac{\pmh(k,\gnl)}{\pmh(k,0)}-1 &= 
\frac{\Delta b(k,\gnl)}{b(M)}+2\frac{\delta\sigma_8}{\sigma_8} 
\label{eq:ratios} \\
\frac{\phh(k,\gnl)}{\phh(k,0)}-1 &= 
\left(1+\frac{\Delta b(k,\gnl)}{b(M)}\right)^2 
+2\frac{\delta\sigma_8}{\sigma_8} - 1 \nonumber\;,
\end{align}
where $\Delta b(k,\gnl)$ is generally the sum of a scale-dependent and
a scale-independent term.  One should bear in mind that the
scale-independent shift  $2\delta\sigma_8/\sigma_8$ arises from the
matter power spectrum and, therefore, is distinct from the term
$-2\nu\delta\sigma_8/\sigma$ appearing in $\Delta b_{\rm I}$.
Following \cite{2009MNRAS.396...85D}, we shall also quantify the
departure from the theoretical scaling as a function of wavemode
amplitude with the ratio $\Delta b^{s}/\Delta b^{t}$, where  $\Delta
b^{s}$ is the non-Gaussian bias correction measured from the
simulation and $\Delta b^{t}$ is Eq.~(\ref{eq:dbiaseff}).

We interpolate the dark matter particles and halo centres onto a
regular cubical mesh. The  resulting dark matter and halo fluctuation
fields are then Fourier transformed to yield the matter-matter,
halo-matter and halo-halo power spectra $\pmm(k)$, $\pmh(k)$ and
$\phh(k)$, respectively. These power spectra are measured for a range
of halo masses and redshifts, covering the relevant  range of
statistical properties corresponding to the available data sets of
galaxy or quasar populations with different luminosities and
bias. Note that these quantities are computed on a 512$^3$ grid, whose
Nyquist wavenumber is sufficiently large ($\approx 1\hmmpc$) to allow
for an accurate measurement of the power in wavemodes of amplitude
$k\lesssim 0.1\hmmpc$.  The halo power spectrum is corrected for the
shot-noise due to the discrete nature of dark matter haloes, which we
assume to be the standard Poisson term $1/\nh$.  This discreteness
correction is negligible for $\pmm(k)$ due to the large number of dark
matter particles. Yet another important quantity is the linear halo
bias $b(M)$ which must be measured accurately from the Gaussian
simulations as it controls the magnitude of the scale-dependent
shift. Here, we shall use the ratio $\pmh(k)/\pmm(k)$ as a proxy for
the halo bias since it is less  sensitive to shot-noise.

\begin{figure}
\center \resizebox{0.5\textwidth}{!}{\includegraphics{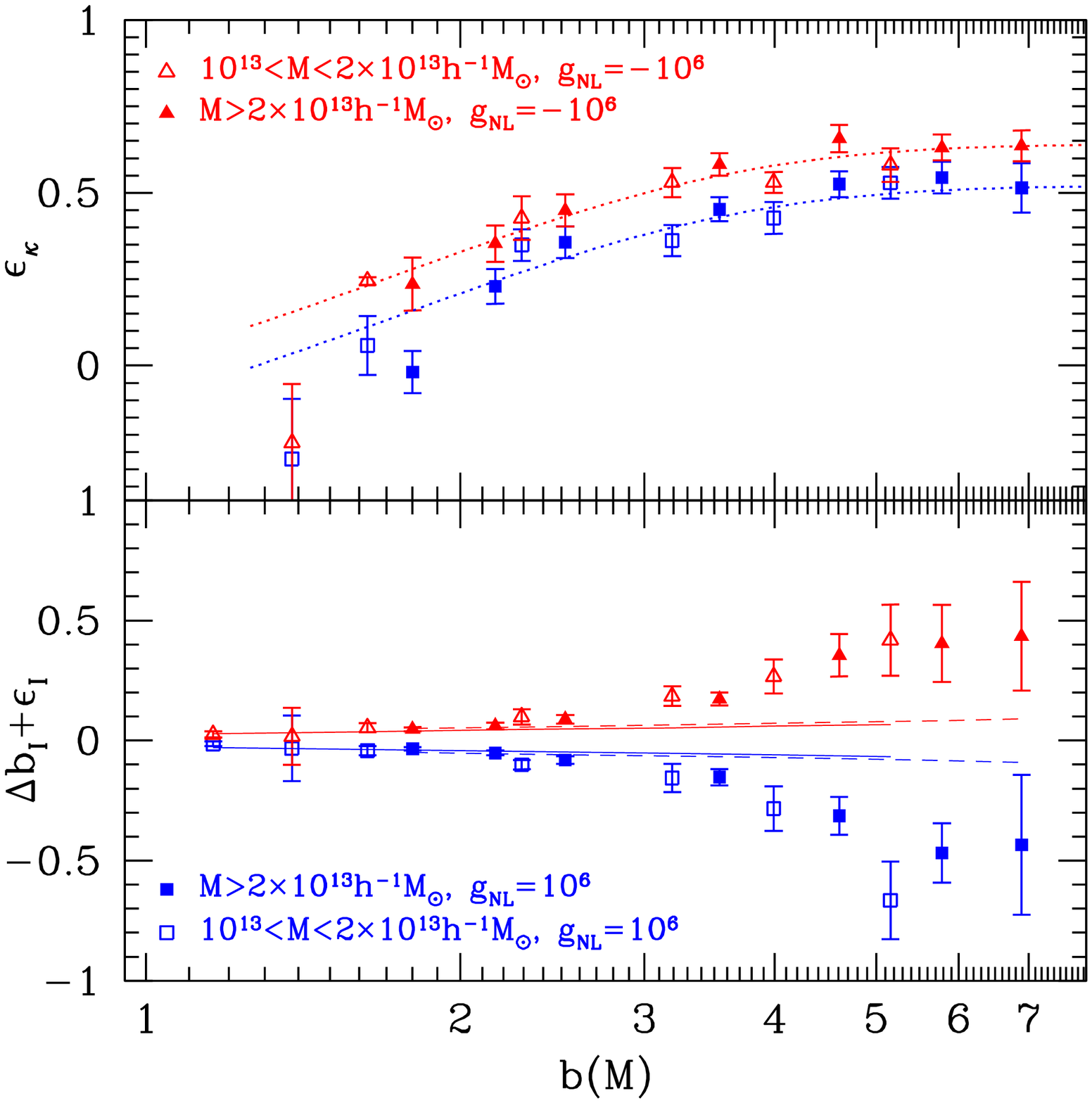}}
\caption{Best-fitting $\ek$ and $\Delta b_{\rm I}+\ei$ as a function
of halo bias and $\gnl=\pm 10^6$, for two different mass cuts as
indicated in the Figure. In the top panel, the dotted curve is our
best fit to $\ek(b,\gnl)$. In the bottom panel, the solid and dashed
lines show the scale-independent shift $\Delta b_{\rm I}$ predicted by
a peak-background split, Eq.~(\ref{eq:gshifti}), for the low and high
mass samples, respectively. }
\label{fig:pfit}
\end{figure}

\begin{figure*}
\center \resizebox{0.8\textwidth}{!}{\includegraphics{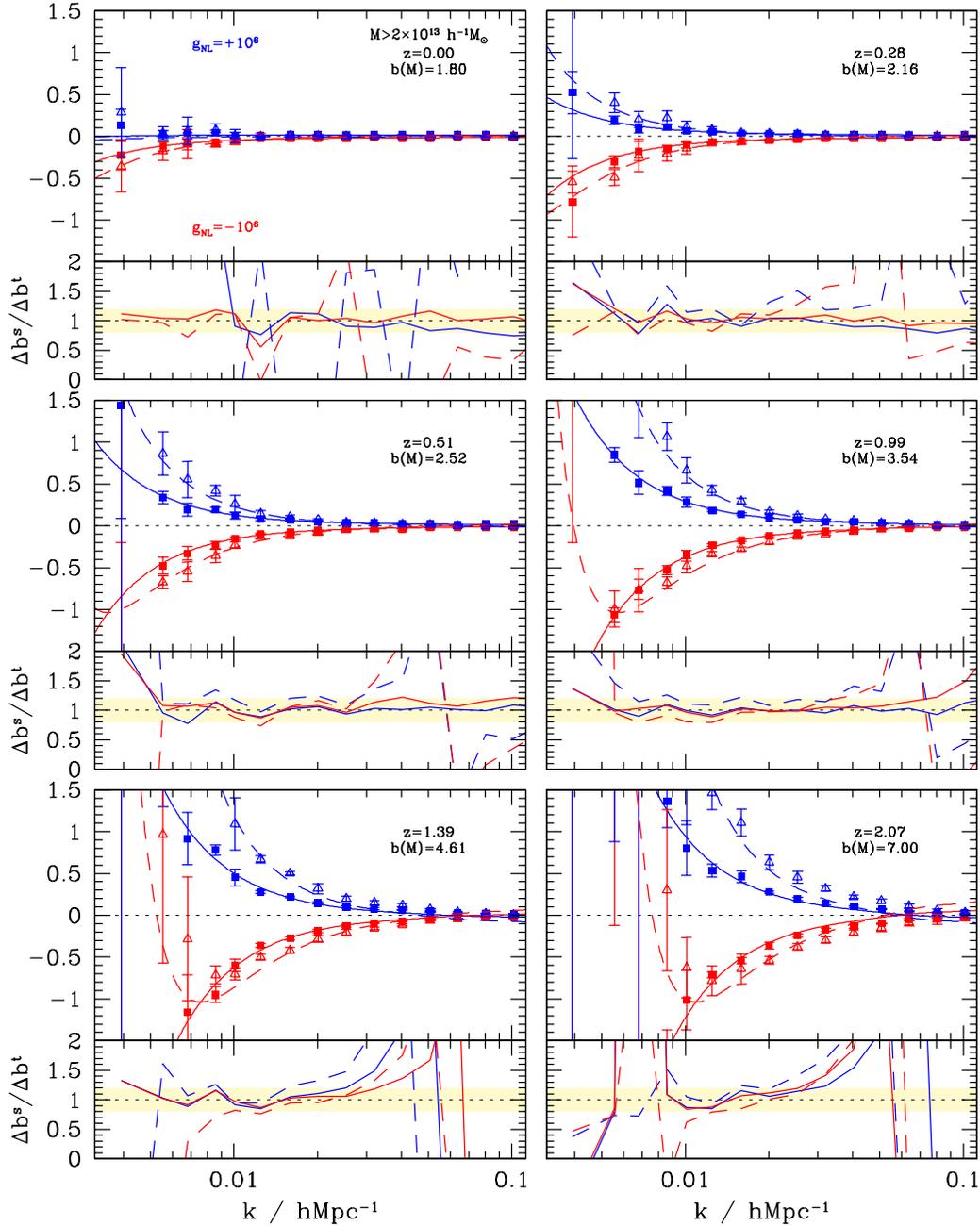}}
\caption{Non-Gaussian bias correction measured in the simulation
outputs at redshift $0<z<2$ for haloes of mass $M>2\times
10^{13}\mdh$. In each panel, the upper plot shows the ratio
$\phh(k,\gnl)/\phh(k,0)-1$ (dashed curves, empty symbols) and
$\pmh(k,\gnl)/\pmh(k,0)-1$ (solid curves,  filled symbols). The error
bars represent the scatter among 5  realisations. The respective
output redshift and linear halo bias are  also quoted. The bottom of
each panel displays the departure from the theoretical prediction,
$\Delta b^s/\Delta b^t$. The shaded area indicates the domain where
the deviation is less than 20 per cent. The parameters $\ek$ and $\ei$
are fitted individually to each sample. For illustration, $\ek$ takes
the best-fit values 0.06 and 0.60 for the lowest and highest biased
samples, respectively.}
\label{fig:bgnl1}
\end{figure*}

\begin{figure*}
\center \resizebox{0.8\textwidth}{!}{\includegraphics{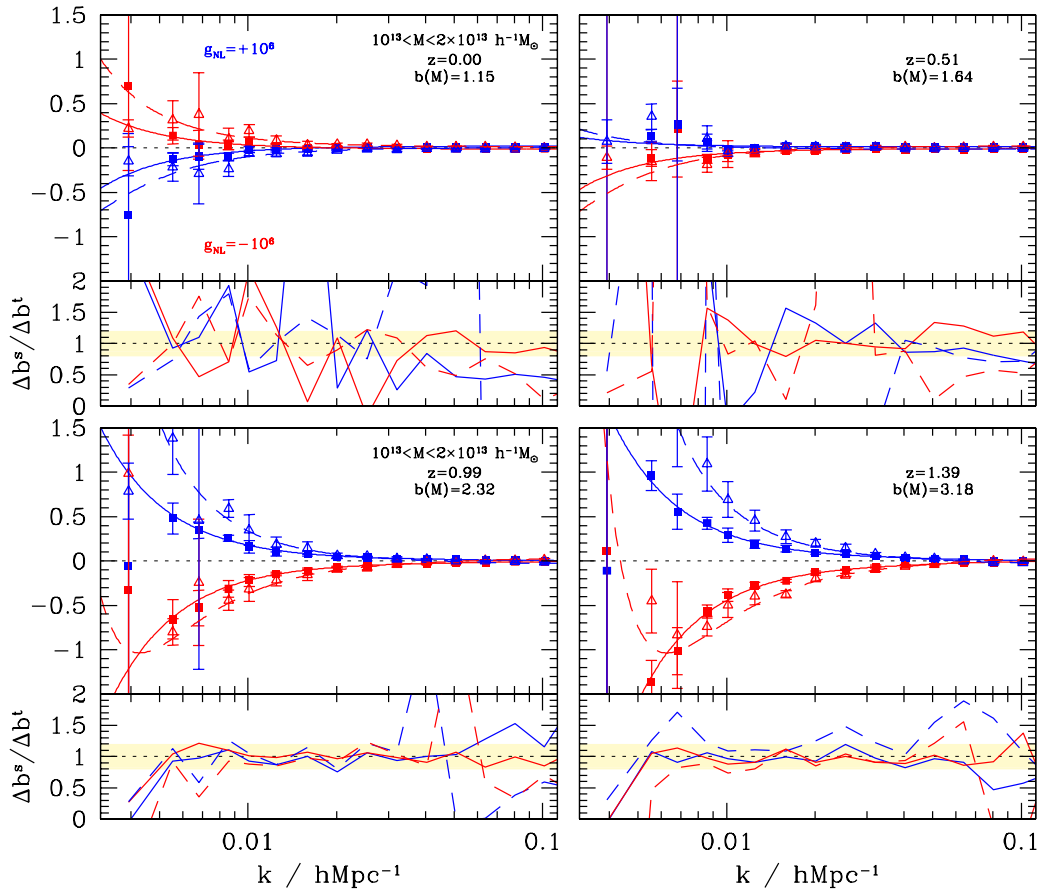}}
\caption{Same as Fig.~\ref{fig:bgnl1} but for haloes in the mass range
$10^{13}<M<2\times 10^{13}\mdh$.}
\label{fig:bgnl2}
\end{figure*}

\subsubsection{An effective non-Gaussian bias correction}

Summarizing the analytical considerations of
Sec. \ref{sub:biastheory}, non-Gaussianity of the $\gnl$ type add a
correction $\Delta b(k,\gnl)$  to the bias $b(k)$ of dark matter
haloes which is at leading order
\begin{equation}
\Delta b(k,\gnl)=\Delta b_\kappa(k,\gnl)+\Delta b_{\rm I}(\gnl)\;,
\label{eq:dbias}
\end{equation}
We found, however, that this theoretical expectation significantly
overestimates the magnitude of the non-Gaussian bias shift measured
from the simulations. This is exemplified in  the top panel of
Fig.~\ref{fig:dbias}, where $\pmh(k,\gnl)/\pmh(k,0)-1$ is plotted for
haloes of mass $M>2\times 10^{13}\mdh$ identified at $z=0.5$. Clearly,
the predicted scale-dependent correction $\Delta b_\kappa$ is much
steeper than measured from the halo samples. In order to improve the
agreement with the numerical data, we modify the above relation as
follows~:
\begin{equation}
\Delta b(k,\gnl) = \ek\,\Delta b_\kappa(k,\gnl) +  \bigl[\Delta b_{\rm
I}(\gnl) + \ei \bigr]\;,
\label{eq:dbiaseff}
\end{equation}
and treat $\ek$ and $\ei$ as free parameters that we fit to our
measurements of the cross-power spectrum (weighted by the scatter
among 5 realisations) in the range $0.005\leq k\leq 0.03\hmmpc$ where
the scale-dependent effect is largest. The bottom panel of
Fig.~\ref{fig:dbias} shows the resulting best-fit contributions
$\ek\Delta b_\kappa$ and $\Delta b_{\rm I}+\ei$ for the halo sample
mentioned above.  As seen in Fig.~\ref{fig:pfit}, $\ek$ and $\ei$
appear to  depend mainly upon the linear halo bias $b(M)$ and the
coupling parameter $\gnl$, although dependencies on redshift or other
halo observables are not excluded (The data is too noisy for a
reliable estimate of these). The most striking feature of
Fig.~\ref{fig:pfit} is the functional dependence of $\ek$ on $b(M)$
and $\gnl$. Firstly, $\ek$ is a monotonically increasing function of
the bias and never reaches unity, even for the most biased samples for
which the high peak approximation should be valid. Secondly, $\ek$ is
noticeably larger for $\gnl=-10^6$, suggesting thereby that second
(and higher) order contributions to the scale-dependent bias may be
important. Furthermore, for $b\lesssim 1.5$ where the high peak
approximation breaks down, there is some evidence that the effect
reverses sign. The bottom panel of Fig.~\ref{fig:pfit} shows that the
scale-independent correction has sign opposite to that of $\gnl$, in
agreement with theoretical expectations from the peak-background split
(see Sec.\ref{sub:biastheory}). However, whereas for $b\lesssim 3$ the
magnitude of the  correction is comparable to that predicted by
Eq.~(\ref{eq:gshifti}), it is considerably larger for $b\gtrsim 3$,
reaching up to 5-10 per cent of the linear halo bias.

Assuming $\ek$ is a function of $b(M)$ and $\gnl$ only and asymptotes
to a constant in the highly biased limit, we find that the following
parametrised form
\begin{equation}
\ek(b,\gnl)=c_1-c_2\,\gnl-\frac{c_3}{1+\exp(c_4 b)}
\label{eq:fit}
\end{equation}
captures reasonably well the increase of $\ek$ with halo bias for
$1.5<b(M)<7$.  The best-fit values of the parameters are
\begin{gather}
c_1=0.59\pm 0.03, \quad c_2=(6.0\pm 0.9)\times 10^{-8} \\
c_3=2.1\pm 0.5, \quad c_4=0.88\pm 0.13 \nonumber\;. 
\end{gather}
We do not provide a fitting formula for $\ei$ (or $\Delta b_{\rm
I}+\ei$) since it is not directly measurable in real data.

Since both the kurtosis of the initial density field and the mass
function of dark matter haloes are in fairly good agreement with
theoretical  expectations, the discrepancy in the scale-dependent bias
$\Delta b_\kappa(k,\gnl)$ indicates that it is the high peak
approximation considered here which is inaccurate, even in the limit
$b(M)\gg 1$ where it is supposed to work best. Our perturbative
approach may be one of the reasons for this mismatch. Namely, we have
derived the scale-dependent bias $\Delta b_\kappa(k,\gnl)$ under the
assumption that the non-Gaussian correction to the halo power spectrum
is small whereas, for most of the  bias range covered by our halo
catalogues, the effect is in fact already very  large at $k\lesssim
0.01\hmmpc$.  Note, however, that for the quadratic coupling
$\fnl\phi^2$, this perturbative treatment predicts an effect of the
right magnitude \cite{2009MNRAS.396...85D}.  Explaining these findings
clearly requires a better theoretical understanding, which we leave
for a future investigation. Notwithstanding this, the phenomenological
prescription (\ref{eq:dbiaseff}) with parameters $\ek$ and $\ei$
fitted to the data provides, as we will see below, a good  description
of the large-scale halo power spectrum in simulations of $\gnl$ models.

\subsubsection{Non-Gaussian bias from auto- and cross-power spectra}

We have measured auto- and cross-power spectra for a range of halo
masses and redshifts spanning the range $0<z<2$.  The ratios defined
in Eq.(\ref{eq:ratios}) are shown in Figs \ref{fig:bgnl1} and
\ref{fig:bgnl2}  as a function of wavenumber for the mass threshold
$M> 2\times 10^{13}\mdh$ and the mass bin $10^{13}<M<2\times
10^{13}\mdh$, respectively.  The fractional deviation $\Delta
b^{s}/\Delta b^{t}$ is also shown at the bottom of each panel. The
shaded region indicates a departure less than 20 per cent. Error bars
denote the scatter around the mean and, therefore, may underestimate
the true errors as they are computed from a small number of
realisations. Note that, in order to reduce  the impact of sampling
variance, we first compute the ratios $\pmh(k,\gnl)/\pmh(k,0)$ and
$\phh(k,\gnl)/\phh(k,0)$ for each realisation before calculating the
average.

As we can see, once $\ek$ and $\ei$ are fitted to the ratio of
cross-power spectra, the theoretical prediction
Eq.~(\ref{eq:dbiaseff}) provides a reasonable description of the
non-Gaussian bias in the halo power spectrum $\phh$, indicating that
non-Gaussianity does not generate much stochasticity and the predicted
scaling $\Delta b_\kappa(k,\gnl) \propto k^{-2}T(k)^{-1}$ applies
equally well for the auto- and cross-power spectrum. This was also
found to be true in $\fnl$ models \cite{2009MNRAS.396...85D}. The
inclusion of a scale-independent correction $\Delta b_{\rm I}+\ei$
significantly improve the agreement at $k\lesssim 0.03\hmmpc$. For the
highly biased samples $b>4$ however, this correction is so large that
the non-Gaussian bias shift reverses sign at wavenumber $k\gtrsim
0.05\hmmpc$. Such an effect is not seen in the simulations, but we
expect large deviations from the relation (\ref{eq:dbiaseff}) in that
range of wavenumber, where second- and higher-order corrections
induced by the cubic coupling $\gnl\phi^3$ together with the nonlinear
bias created by the gravitational evolution of matter density
fluctuations may become important. Even though the data is noisier due
to the low number density of haloes, it is worth noting that, for the
highly biased samples at $k\lesssim 0.01\hmmpc$, the cross-power
spectrum $\pmh(k,\gnl=-10^6)$ goes negative while $\phh(k,\gnl=-10^6)$
remains positive and increases sharply, in agreement with the analytic
prediction. Still, there is some evidence that the ratio
$\phh(k,\gnl=-10^6)/\phh(k,0)-1$ saturates at a value noticeably
larger than -1 before the sharp upturn, whereas our model predicts
$2\delta\sigma_8/\sigma_8-1\approx -0.96$ at the minimum.

Fig.~\ref{fig:bgnl2} further explore the effect in the low mass
samples, for which the $z=0$ haloes with $b(M)\approx 1.15$ constitute
an almost unbiased sample of the density field. In this case, the sign
of the scale-dependent contribution is reversed, namely, the
large-scale halo power spectrum in simulations of $\gnl=-10^6$ is
enhanced relative to that of the Gaussian ones. This is in rough
agreement with the theory, which predicts a similar effect for
$b(M)<1$. Again, haloes with a similar bias also have a comparable
scale-dependent and scale-independent bias regardless of mass or
redshift. Finally, note that the sample at $z=1.39$ shown in
Fig.~\ref{fig:bgnl2} corresponds closely to the quasar sample used by
\cite{2008JCAP...08..031S}, for which $z=1.8$ and $b=2.7$.

\section{Constraints on the coupling parameter $\gnl$}
\label{sec:bound}

\subsection{Constraints on $\gnl$ from current large-scale structure data}

Reference \cite{2008JCAP...08..031S} took advantage of the
scale-dependence of the bias to constrain $\fnl$ from a sample of
highly biased luminous red galaxies (LRGs) and quasars (QSOs). It is
straightforward to translate their 2-$\sigma$ limit $-29 < \fnl < +69$
into a constraint on $\gnl$ since the non-Gaussian scale-dependent
bias $\Delta b_\kappa(k,\gnl)$ has  the same functional form as
$\Delta b_\kappa(k,\fnl)$.

Constraints will arise mostly from the QSO sample at median redshift
$z=1.8$, which covers a large comoving volume and is highly biased,
$b=2.7$.  In light of our results (see Fig.~\ref{fig:pfit}), we expect
the parameter  $\ek(b,\gnl)$  to vary with $\gnl$. However, in order
to simplify the analysis, we will assume that, at fixed $b$,
$\ek(b,\gnl)$ is given by the mean of $\ek(b,\gnl=\pm 10^6)$. For a
sample with bias $b\sim 2.7$,  this implies $\ek\simeq
0.4$. Furthermore, assuming the typical mass of QSO-hosting
haloes is $\sim 10^{13}\mdh$ yields $S_3^{(1)}\!(M)\simeq 2.3\times
10^{-4}$. Hence, the multiplicative factor $(1/4)\,\dc(z)\ek
S_3^{(1)}\!(M)$ is approximately  $\simeq 8.4\times 10^{-5}$. Our
limits on $\gnl$ thus are
\begin{equation}
-3.5\times 10^5 < \gnl < +8.2 \times 10^5
\end{equation}
at 95\% confidence level.  The scale-independent correction $\Delta
b_{\rm I}+\ei$ is not directly measured as it adds up to the bias $b$
which is fitted to the data. For the limits obtained here, $|\Delta
b_{\rm I}+\ei|$ should be much smaller than $b$ and can thus be
ignored.  Note also that, whereas the non-Gaussian bias scales as
$D(z)^{-1}$ in $\fnl$ models, we have $\Delta b(k,\gnl)\propto
D(z)^{-2}$ for $\gnl$ non-Gaussianity, so one can achieve relatively
larger gains from measurements of high redshift tracers.  In fact, the
extent to which one can improve the observational bounds will strongly
depend on our ability to minimize the impact of sampling variance
caused by the random nature of the wavemodes, and the shot-noise
caused by the discrete nature of the tracers. By comparing differently
biased tracers of the same surveyed volume \cite{2009PhRvL.102b1302S}
and suitably weighting galaxies (e.g. by the mass of their host halo)
\cite{2009JCAP...03..004S,2009PhRvL.103i1303S}, it should be possible
to circumvent these problems and considerably improve the detection
level.

\subsection{Predictions for future LSS surveys} 

References
\cite{2008ApJ...684L...1C,2008PhRvD..78l3519M,2009PhRvL.102b1302S,
2008arXiv0810.0323M} applied the Fisher matrix formalism to
forecast constraints on $\fnl$ from forthcoming galaxy redshift
surveys. Here, we will simply try to estimate the detection limit 
for $\gnl$. Following
\cite{2008PhRvD..78l3519M,2008ApJ...684L...1C}, we consider a (nearly
spherical) survey of volume $V$.  Assuming the Fourier modes are still
uncorrelated and Gaussian distributed, the total signal-to-noise
squared reads
\begin{equation}
\left(\frac{S}{N}\right)^2\approx \frac{V}{4\pi^2}
\int_{k_{\rm min}}^{k_{\rm max}}\!\! dk\, k^2 
\left[\left(1+\frac{\Delta b_\kappa}{b}\right)^2-1\right]^2
\label{eq:snsurvey}
\end{equation}
in the limit where sampling variance dominates the errors. Here,
$k_{\rm min}\sim \pi/V^{1/3}$ is the smallest wavemode accessible and
$k_{\rm max}$ is not necessarily finite since the integral does
converge as one takes $k_{\rm max}$ to infinity. Substituting the
expression Eq.~(\ref{eq:gshiftk}) for the scale-dependent bias $\Delta
b_\kappa(k,\gnl)$ and setting $T(k)\equiv 1$ over the wavenumber range
across which the integral is performed, we arrive at
\begin{equation}
\left(\frac{S}{N}\right)^2\approx\frac{V}{\pi^2} (k_\star^2)^2
\left(\frac{1}{k_{\rm min}}-\frac{1}{k_{\rm max}}\right)\;,
\end{equation}
where
\begin{equation}
k_\star^2\simeq 5.0\times 10^{-12}\gnl\ek\frac{(1-1/b)}{D^2(z)}
\left(\frac{S_3^{(1)}}{10^{-4}}\right)\hhmmpc\;.
\end{equation}
We have also assumed $|\gnl|\lesssim 10^5$, such that $|k_\star^2|$ is at
most  of the order of $k_{\rm min}^2$ and the term linear in $\Delta
b_\kappa/b$ dominates the signal. When $k_{\rm min}\ll k_{\rm max}$,
we can further simplify $(S/N)^2$ to
\begin{align}
\left(\frac{S}{N}\right)^2 &\approx 8.1\times
10^{-13}\gnl^2\ek^2\,\left(1-\frac{1}{b}\right)^2 D(z)^{-4} \nonumber \\ 
& \quad\times 
\left(\frac{S_3^{(1)}}{10^{-4}}\right)^2\left(\frac{V}{\hhhgpc}\right)^{4/3}
\;.
\end{align}
Note the strong sensitivity of the signal-to-noise squared to the
growth factor  $D(z)$ (For $\fnl$ non-Gaussianity, this dependence is
only $D(z)^{-2}$).

To highlight the improvement one could achieve with future galaxy
surveys,  it is useful to first calculate the detection limit for the
SDSS LRG sample centred at $z\sim 0.3$ and covering a volume
$v\approx 2\hhhgpc$.  Assuming a linear bias $b=2$ and a skewness
parameter $S_3^{(1)}\sim 2\times 10^{-4}$ appropriate for haloes of
mass $M\sim 10^{12}-10^{13}\mdh$, the minimum $\gnl$ detectable at the
1-$\sigma$ level is  $\simeq 10^6$  for a correction factor $\ek=0.3$
which we read off from Fig.~\ref{fig:pfit}. For a survey configuration
analogous to SDSS-III/BOSS~\footnote{www.sdss3.org}, with central
redshift $z=0.5$ and a comoving volume $V=6\hhhgpc$, the minimum
$\gnl$ would be   $\sim 4\times 10^5$ for galaxies tracing haloes of
similar mass and bias.  Finally, for a configuration like EUCLID 
~\footnote{http://sci.esa.int/science-e/www/object/index.cfm?fobjectid=42266}
with a $V=100\hhhgpc$ survey  centred at $z=1.4$, the detection limit
would be $\sim 2.1\times 10^4$. Clearly, these limits are only
indicative: they may be significantly improved by selecting  highly
biased, high redshift (single- or multi-)tracers. Nevertheless, this
shows that future galaxy surveys should furnish interesting
constraints on the size of the cubic coupling $\gnl\phi^3$.

\subsection{Predictions for CMB temperature anisotropies}

The CMB trispectrum provides an alternative probe of local,
non-quadratic correction to the Gaussian curvature perturbations, so
it is  interesting to assess the sensitivity of this statistics to the
nonlinear parameter $\gnl$.

The temperature anisotropy field is conveniently decomposed into
spherical harmonics, $\Delta T(\nn)/T=\sum_{l m} a_l^m Y_l^m(\nn)$.
As shown in \cite{2001PhRvD..64h3005H,2002PhRvD..66f3008O},
statistical isotropy and invariance under parity transformation
$\nn\rightarrow -\nn$ implies that the  4-point correlation of the
spherical harmonic coefficients $a_l^m$ takes the form
\begin{align}
\label{eq:cmbfourpt}
\la a_{l_1}^{m_1}a_{l_2}^{m_2}a_{l_3}^{m_3}a_{l_4}^{m_4}\ra &=
\sum_{LM} (-1)^M \left(\begin{array}{ccc} l_1 & l_2 & L \\ m_1 & m_2 &
-M \end{array}\right) \\ & \quad \times \left(\begin{array}{ccc} l_3
& l_4 & L \\ m_3 & m_4 & M \end{array}\right) Q_{l_3 l_4}^{l_1 l_2}(L)
\nonumber \;.
\end{align}
Here, $Q_{l_3 l_4}^{l_1l_2}(L)$ is the angular average trispectrum and
brackets are Wigner-3j symbols. Statistical homogeneity also implies
that $Q_{l_3 l_4}^{l_1l_2}(L)$  is independent of position. The
connected part of the trispectrum, $T_{l_3 l_4}^{l_1l_2}(L)$, encodes
information about non-Gaussianity and is obtained by subtracting a
Gaussian piece constructed from the power spectra $C_l$. 
Eq.(\ref{eq:cmbfourpt}) can be inverted with
the aid of the orthogonality  of the Wigner-3j symbols to form an
estimator for the CMB trispectrum.

The signal-to-noise for the CMB trispectrum $T_{l_3 l_4}^{l_1 l_2}(L)$
summed up to a certain $l_{\rm max}$ is \cite{2002PhRvD..66f3008O}
\begin{equation}
\left(\frac{S}{N}\right)^2\!\!(<l_{\rm max})\approx 
\sum_{l_1>l_2>l_3>l_4}^{l_{\rm max}}
\sum_L \frac{|T_{l_3 l_4}^{l_1 l_2}(L)|^2}{(2L+1)
C_{l_1}C_{l_2}C_{l_3} C_{l_4}}
\label{eq:sntri}
\end{equation}
when cosmic variance dominates the errors. Otherwise,  one shall include
a contribution from the power spectrum of the detector  noise to the
$C_l$. Galactic foreground subtraction on a fraction $1-f_{\rm sky}$
would further reduce $(S/N)^2$ by a factor of $f_{\rm sky}$.

Neglecting the ISW effect, the Sachs-Wolfe provides a useful
order-of-magnitude estimate of the signal-to-noise as long as $l_{\rm
max}$ does not exceed $\lesssim 100$
\cite{1994ApJ...430..447G,2001PhRvD..63f3002K,2002PhRvD..66f3008O,
2006PhRvD..73h3007K}. The calculation is performed in Appendix 
\ref{app:noisetricmb}. We find that the signal-to-noise can be recast  
into the compact form
\begin{align}
\left(\frac{S}{N}\right)^2\!\!(< l_{\rm max}) &= \frac{9}{2}\gnl^2 
A_\phi^2\left\{\frac{1}{6}\int_{-1}^{+1}\!\! dx\, s_{l_{\rm max}}^3(x)
t_{l_{\rm max}}(x)\right. \nonumber \\
& \qquad + \left.\frac{1}{2}\int_{-1}^{+1}\!\! dx\, r_{l_{\rm max}}^2(x)
s_{l_{\rm max}}^2(x)\right\}
\label{eq:snsw}
\end{align}
where the auxiliary functions $r_l(x)$, $s_l(x)$ and $t_l(x)$ are
defined as
\begin{align}
r_l(x) &= \sum_{k=2}^l (2k+1) P_k(x) \\
s_l(x) &= \sum_{k=2}^l \frac{(2k+1)}{k(k+1)}P_k(x) \\
t_l(x) &= \sum_{k=2}^l (2k+1)k(k+1) P_k(x) \;.
\end{align}
Here, $P_l(x)$ are Legendre polynomials. Note that we have excluded
the monopole and dipole from the summation since these modes are
unobservable. We have also assumed a nearly scale-invariant spectrum
$n_s\approx 1$.

\begin{figure}
\center \resizebox{0.5\textwidth}{!}{\includegraphics{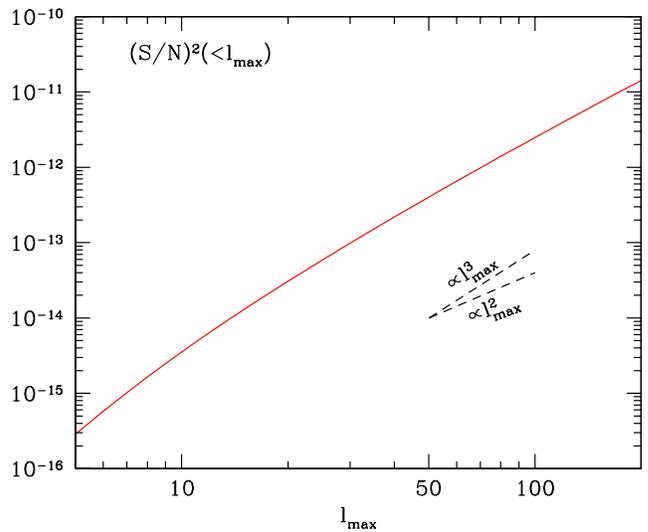}}
\caption{Signal-to-noise ratio squared for the CMB trispectrum as a
function of the maximum multipole $l_{\rm max}$. We have assumed
$\gnl=1$ and $f_{\rm sky}=1$. }
\label{fig:sncmb}
\end{figure}

Figure.~\ref{fig:sncmb} shows the signal-to-noise ratio squared in the
Sachs-Wolfe approximation for the CMB trispectrum assuming $\gnl=1$.
Although this approximation breaks down for  $l_{\rm max}\gtrsim 100$,
we have extended the calculation up to  $l_{\rm max}=200$ so as to
extrapolate more robustly the $l_{\rm max}$-dependence to small
angular resolution.  A power-law fit to $(S/N)^2$ in the range $50\leq
l_{\rm max}\leq 200$ gives
\begin{equation}
\left(\frac{S}{N}\right)^2\!\!(<l_{\rm max})\simeq 
2.43\times 10^{-17}\gnl^2 
\left(\frac{A_\phi}{10^{-9}}\right)^2l_{\rm max}^{2.6}\;.
\label{eq:sncmbscaling}
\end{equation}  
Our results appear consistent with the findings of
\cite{2002PhRvD..66f3008O} shown in their Fig.2. However, our constant
of proportionality is about 20-30 times larger,  presumably because
they  adopted a lower fluctuation amplitude (compare also their
prediction for the $\fnl$ model with that of
\cite{2006PhRvD..73h3007K}).  Adding the information encoded
in  temperature-polarization trispectra may enhance $(S/N)^2$ by a
factor of a few \cite{2001PhRvD..64h3005H}.

Assuming the scaling Eq.(\ref{eq:sncmbscaling}) persists well beyond
the range over which the Sachs-Wolfe effect dominates, the minimum
$\gnl$  detectable at 1-$\sigma$ level is $\gnl\simeq 20$, 7.9, 3.2,
1.9 and $1.3\times 10^4$ for $l_{\rm max}=250$, 500, 1000, 1500 and
2000.  A more realistic calculation should include the full radiation
transfer function, detector noise etc. In this respect, detailed
calculations have shown that, for the quadratic coupling $\fnl\phi^2$,
$(S/N)^2$ of the CMB bispectrum and trispectrum closely follows the
behaviour obtained in the Sachs-Wolfe approximation
\cite{2002PhRvD..66f3008O,2006PhRvD..73h3007K}. It seems reasonable,
then, to expect that this is also true for $\gnl\phi^3$.

While our predictions are qualitative they show that, for the WMAP CMB
temperature measurement  ~\footnote{http://map.gsfc.nasa.gov/} (which
we approximate as a noise-free experiment with $l_{\rm max}\sim 250$),
no detection of a significant trispectrum implies $|\gnl|\leq 2\times
10^5$ at the 1-$\sigma$ level. This is of the same order as the limit
we derived from the QSO sample analyzed by
\cite{2008JCAP...08..031S}. For a PLANCK-like experiment
\footnote{http://sci.esa.int/science-e/www/area/index.cfm?fareaid=17}
($l_{\rm max}\sim 1500$), no evidence for a trispectrum would imply
$|\gnl|\leq 1.3\times 10^4$ at the 1-$\sigma$ level. This is
comparable to the detection limit that could be achieved with an
all-sky survey such as EUCLID.

\section{Effect of non-Gaussianity with non-zero $\fnl$ and $\gnl$}
\label{sec:biasfnlgnl}

In this Section, we examine the halo multiplicity function and
large-scale bias in numerical simulations of structure formation with
non-zero coupling parameters $(\fnl,\gnl)=(\pm 100,-3\times 10^5)$. We
show that the results are consistent with those obtained from the
simulations with non-vanishing $\gnl$ solely.

\subsection{Mass function}

\begin{figure}
\center \resizebox{0.5\textwidth}{!}{\includegraphics{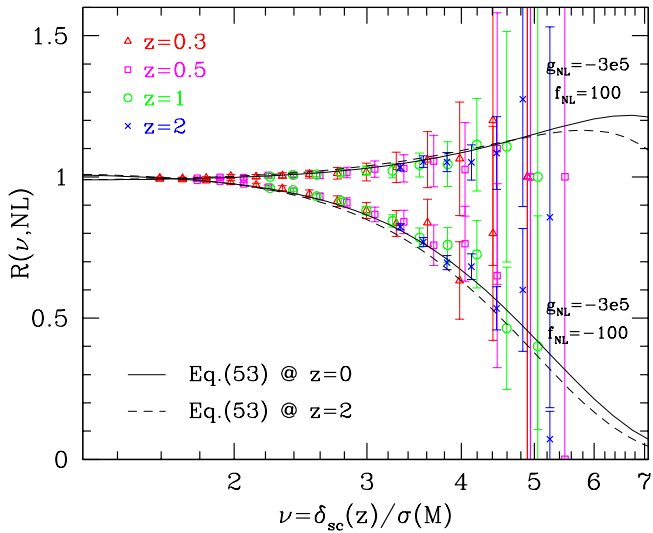}}
\caption{Fractional correction to the Gaussian multiplicity function
of dark matter haloes as a function of the peak height $\nu(M,z)$ for
$\fnl=\pm 100$ and a cubic coupling parameter $\gnl=-3\times 10^5$.
The solid and dashed curves show the theoretical prediction
Eq.(\ref{eq:fnufg})  at $z=0$ and 2, respectively. Error bars denote
Poisson errors.}
\label{fig:fnufg}
\end{figure}

It is straightforward to calculate the fractional deviation from the
Gaussian mass function,  Eq.(\ref{eq:thiswork}), to non-zero $\fnl$
and $\gnl$. Again, we start with the MVJ formula and neglect second
order corrections such as $(\sigma S_3)^2$ etc. Adjusting the
coefficient of the terms $\nu\sigma S_3$ and $\nu^2\sigma^2 S_4$ to
that of  the small $\nu$ expansion obtained by
\cite{2008JCAP...04..014L}, we arrive at
\begin{align}
R(\nu,{\rm NL}) &= \exp\left[\frac{\nu^3}{3!}\sigma
S_3+\frac{\nu^4}{4!}\sigma^2 S_4 +\nu^2\delta\sigma_8\right]
\nonumber \label{eq:fnufg}
\\ & \quad\times\left\{1-\frac{\nu}{2}\sigma S_3-\frac{\nu}{6}
\frac{d(\sigma S_3)}{d\ln\nu}\right. \\ &
\qquad\quad\left. -\frac{\nu^2}{4}\sigma^2 S_4-\frac{\nu^2}{4!}
\frac{d(\sigma^2 S_4)}{d\ln\nu}\right\}\nonumber\;,
\end{align}
where the shorthand notation ${\rm NL}$ designates the combination
$(\fnl,\gnl)$.  In Fig.~\ref{fig:fnufg}, this theoretical prediction
is compared  $R(\nu,{\rm NL})$ measured  in non-Gaussian simulations
of $(\fnl,\gnl)=(\pm 100,-3\times 10^5)$. We account for the fact that
the amplitude of density fluctuations is renormalised by
$\delta\sigma_8\approx 0.0045$. Fig.~\ref{fig:fnufg} demonstrates that
our approximation is in good agreement with the data, although it
slightly overestimates the effect at $z=2$ when $\fnl=-100$. For
$\fnl=100$, the positive and negative contributions from the quadratic
and cubic coupling,  respectively, almost cancel each other and
flatten the fractional deviation over most of the mass range probed by
the simulations.

The scale-independent bias shift which arises from the change in the
mean number density of haloes can again be estimated using the 
peak-background split. We find
\begin{align}
  \Delta b_{\rm I}({\rm NL}) &\approx
  -\frac{1}{\sigma}\biggl[\frac{1}{3!}  \left(\nu^3-3\nu\right)\sigma^2
    S_4 + \frac{1}{2}\left(\nu^2-1\right) \sigma S_3 \biggr. 
    \nonumber \\ 
    & \qquad \biggl. +\frac{1}{4!}\left(\nu^3-8\nu\right)
    \frac{d(\sigma^2 S_4)}{d\ln\nu}-\frac{\nu}{4!}
    \frac{d^2(\sigma^2 S_4)}{d\ln\nu^2}\biggr. \nonumber \\
    & \qquad \biggl. +\frac{1}{6}\left(\nu^2-4\right)
    \frac{d(\sigma S_3)}{d\ln\nu}-\frac{1}{6}
    \frac{d^2(\sigma S_3)}{d\ln\nu^2}\biggr. \nonumber \\
    & \qquad \biggl. + 2\nu\delta\sigma_8 \biggr]
\label{eq:fgshifti}
\end{align}
at the first order in the nonlinear parameters $\fnl$ and $\gnl$.

\subsection{Bias}

Having checked that the amount of non-Gaussianity in the mass function
is also consistent with our simple theoretical expectation when both
$\fnl$ and $\gnl$ are non-zero, we now turn to the clustering of dark
matter haloes.  As shown in Appendix \ref{app:ngbias}, the
non-Gaussian correction to  the halo power spectrum can be written 
down as
\begin{align}
  \Delta\phh(k) &= 4\fnl b_{\rm L}^2\dc(z)\am(k)P_\phi(k) \nonumber \\
  & \quad + 4\fnl^2 b_{\rm L}^2\dc^2(z) P_\phi(k) + 
  \left(\gnl+\frac{4}{3}\fnl^2\right) \nonumber \\
  & \qquad \times b_{\rm L}^2\dc^2(z) S_3^{(1)}\!(M)\am(k) P_\phi(k)\;.
\label{eq:dphhfg}
\end{align}
If we set $\gnl=0$ and keep only the first two terms in the right-hand
side, then the non-Gaussian (Eulerian) halo power spectrum can be cast
into the form
\begin{equation}
\phh(k)=\bigl[ b(M) + \fnl b_\phi(k) \bigr]^2 P_\delta(k)
\end{equation}
where the scale-dependent bias parameter  $b_\phi(k)$ is
\begin{equation}
b_\phi(k)=2 \bigl[b(M)-1\bigr] \dc(z)\,\am^{-1}(k)\;.
\end{equation}
Note that reference \cite{2008PhRvD..78l3519M} obtained this relation
by considering the halo power spectrum implied by a bias relation that
is a local mapping of the density field. In practice, the term
proportional to $P_\phi(k)$ is negligible as it contributes only at
very small wavenumber $k\lesssim 0.001\hmpc$.  The third term in the
right-hand side of Eq.(\ref{eq:dphhfg}) is derived in this paper for the
first time. In the case $\gnl=0$, its magnitude relative to the term
linear in $\fnl$ is $(1/3)\fnl \dc(z) S_3^{(1)}\!(M)$, which is
approximately $\sim 0.03$ at redshift $z=1.8$ and for a mass scale
$M=10^{13}\mdh$.  Although its contribution becomes increasingly
important at higher redshift, it is fairly small for the values of
$\fnl$ considered here. Consequently, we shall employ the approximation
\begin{align}
  \Delta b(k,{\rm NL}) &= \ek\,\Delta b_\kappa(k,\gnl) + 
  \Delta b_\kappa(k,\fnl) \nonumber \\ 
  & \quad + \bigl[\Delta b_{\rm I}({\rm NL})+\ei \bigr]\;.
\label{eq:dbiaseff1}
\end{align}
to describe the non-Gaussian bias of dark matter haloes.

\begin{figure*}
\center \resizebox{0.8\textwidth}{!}{\includegraphics{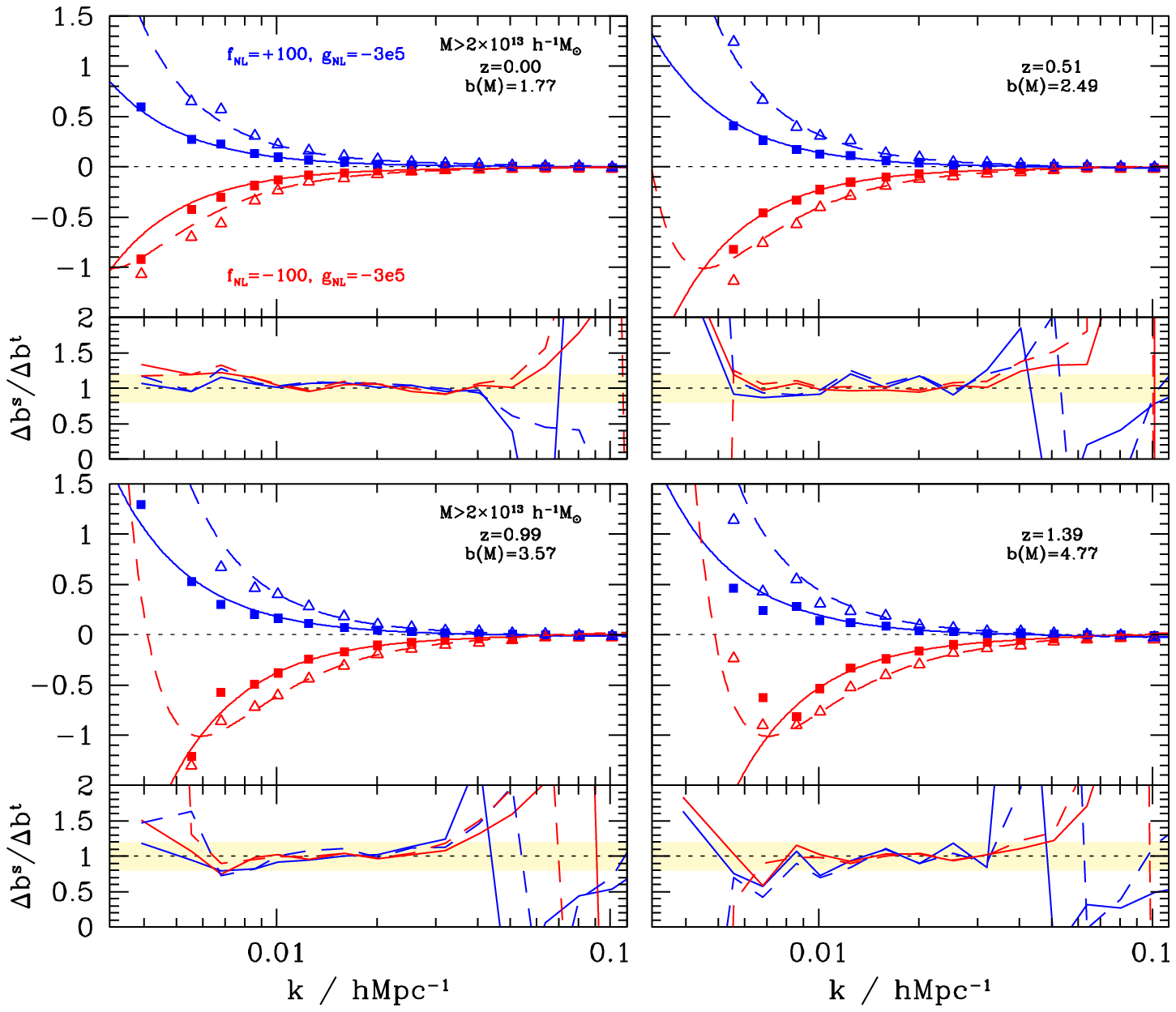}}
\caption{Non-Gaussian bias correction measured in the simulations with
$(\fnl,\gnl)=(\pm 100,-3\times 10 ^5)$ for haloes of mass $M>2\times
10^{13}\mdh$. Error bars are not shown as the data points are
averaged over two realisations solely.}
\label{fig:bgnl3}
\end{figure*}

The quadratic coupling $\fnl\phi^2$ also affect the matter power
spectrum at leading order
\cite{2004PhRvD..69j3513S,2008MNRAS.390..438G},  positive values of
$\fnl$ increasing the small scale power. However, the relative size of
this $k$-dependent correction, $\beta_{\rm m}(k,\fnl)$, is at a per
cent level only in the weakly nonlinear regime $k\lesssim 0.1\hmmpc$
\cite{2008PhRvD..78l3534T,2009MNRAS.396...85D} and fades rapidly as
one goes to larger scales. We will thus neglect it in what follows.

In Fig.~\ref{fig:bgnl3}, the result of measuring ratios of auto- and
cross-power spectra in the simulations with  $(\fnl,\gnl)=(\pm
100,-3\times 10^5)$ is shown at $0<z<1.5$ for the haloes with
$M>2\times 10^{13}\mdh$. We do not quote error bars since the data
points are obtained by averaging over two realisations only.  The
solid and dashed curves show the theoretical prediction
Eq.~(\ref{eq:dbiaseff1}). The value of the multiplicative factor
$\ek(b,\gnl)$ was obtained from the four-parameter formula
Eq.(\ref{eq:fit}), while $\Delta b_{\rm I}+\ei$ was individually
fitted for each halo sample over the wavenumber range
$0.005<k<0.03\hmmpc$. As can be seen, the theoretical expectation
Eq.~(\ref{eq:dbiaseff1}) agrees reasonably well with the numerical
data. This demonstrates that the range of validity of the
non-Gaussian bias formula Eq.~(\ref{eq:dbiaseff}) extends to  smaller
values of $\gnl$ as well as models with non-vanishing $\fnl$ and 
$\gnl$.

The lowest order, $k$-dependent corrections to the Gaussian bias
induced by the quadratic and the cubic coupling are fully degenerated
in the halo power spectrum as they both scale as $\am^{-1}(k)\propto
k^{-2}T(k)^{-1}$.  For the values of $\fnl$ and $\gnl$ and the halo
mass range considered here, the ratio $\Delta b_\kappa(k,\gnl)/\Delta
b_\kappa(k,\fnl)$ increases approximately from  0.25 to 0.5 when the
redshift increases from $z=0$ to 2. It is unclear whether higher order
corrections could help breaking such a degeneracy. A more promising 
alternative may be to measure the bispectrum of dark matter haloes,
which carries much more information about the shape of the primordial
three-point function than the power spectrum of bias tracers
\cite{2009arXiv0904.0497J,2009arXiv0905.0717S}. However, this is 
beyond the scope of this paper.

\section{Discussion}
\label{sec:discussion}

In this paper we explored the effect of a local cubic coupling
$\gnl\phi^3$ on the mass function and bias of dark matter haloes. We
derived analytical expressions for these statistics  and tested them
against the outcome of numerical simulations.

We showed that current theoretical predictions of the non-Gaussian
correction to the mass function reasonably agree with the simulations.
The LV formula \cite{2008JCAP...04..014L} appears to provide  a better
fit to the data than the MVJ formula \cite{2000ApJ...541...10M}, in
agreement with  some of the literature on the subject
\cite{2007MNRAS.376..343K,2008PhRvD..77l3514D}. The two
approximations can be combined to provide an accurate description if
one   adjusts  the low-$\nu$ expansion of the latter so as to match
that of the  former.

We found that the magnitude of the non-Gaussian scale-dependent bias
$\Delta b_\kappa(k,\gnl)$ is suppressed relative  to a theoretical
prediction based on the statistics of highly overdense regions, even
on linear scales $k\lesssim 0.01\hmmpc$.  This suppression is stronger
for  the low biased samples $b\lesssim 3$ and, at fixed value of $b$,
for positive values of $\gnl$. We were able to fit the measured halo
bias at the expense of introducing two free parameters, $\ek$  and
$\ei$, that depend mostly on the halo bias $b(M)$ and the coupling
parameter $\gnl$. These parameters quantify the departure from the
theoretical scale-dependent and scale-independent non-Gaussian bias
correction, respectively. We provide a simple fitting formula for
$\ek(b,\gnl)$, Eq.(\ref{eq:fit}), which should be used when analyzing
observational data. In non-Gaussian simulations of the $\fnl$ type,
the data also hint at a (possibly $\fnl$-dependent) suppression of the
non-Gaussian scale-dependent bias relative to theory  for wavemodes
$k\lesssim 0.03\hmmpc$ \cite{2009MNRAS.396...85D,
2008arXiv0811.4176P,2009arXiv0902.2013G}, but the effect is much
weaker than seen in our simulations of $\gnl$ models. Clearly,  these
results require a better theoretical modelling of the non-Gaussian
halo bias.

Reference \cite{2009arXiv0902.2013G} argued that both the MVJ and LV
analytic formula can be reconciled with measurements of the
non-Gaussian fractional correction to the mass function  once
non-spherical collapse is included. In practice, the critical density
for collapse is replaced by $\dc\rightarrow \sqrt{q}\dc$, where the
value $q=0.75$ is  obtained from a fit to the mass function measured
in simulations \cite{2002MNRAS.329...61S}. Reference
\cite{2009arXiv0903.1251M} claimed that such a relation is a
consequence of the diffusive nature of the critical threshold for
collapse. Their model predicts $q\simeq 0.8$, in good agreement with
the findings of \cite{2009arXiv0902.2013G}. However, we found that
substituting $\dc\rightarrow \sqrt{q}\dc$ in Eq.(\ref{eq:thiswork})
only  modestly improve the agreement with the data.  Regarding the
non-Gaussian bias, it is not obvious how one could justify the
replacement $\dc\rightarrow \sqrt{q}\dc$ given that the linear bias of
our (Gaussian) halo samples converges towards the spherical collapse
prediction $\nu^2/\dc$ for large peak height.

A important ingredient is the choice of the halo identification
algorithm. While we used a spherical overdensity (SO) finder,
reference \cite{2009arXiv0903.1251M} considered a Friends-of-Friends
(FoF) finder with a linking length $b=0.2$. The question of how the
spherical overdensity masses can be mapped onto friends-of-friends
masses remains a matter of debate \cite{2009ApJ...692..217L}.   Clealy
however, since the peak height depends on the halo mass $M$ through
the variance $\sigma(M)$,  any systematic difference will be reflected
in the value of $\nu$ associated to a specific halo sample.  This will
in turn affects the size of the fractional deviation from the Gaussian
mass function at some specified peak height. The sensitivity of the
non-Gaussian mass function and bias to the halo finder will be
presented elsewhere.

The observational bound on  $\fnl$ inferred by
\cite{2008JCAP...08..031S} from the clustering of a high redshift
sample of quasars can be straightforwardly translated into a limit 
on $\gnl$ since $\Delta b_\kappa(k,\gnl)$ also scales as
$k^{-2}T(k)^{-1}$ at low wavenumber. We have obtained
\begin{equation}
-3.5\times 10^5 < \gnl < +8.2 \times 10^5~~ (95\%).
\end{equation}
These are the first limits derived on $\gnl$.  While they are too weak
to provide interesting constraints on  inflationary scenarios such as
the curvaton model, future all-sky redshift surveys should improve
them by a factor  of $\sim 100$. Future CMB observations, including
PLANCK, should also  improve the limits derived here by an order of
magnitude. With these improvements we  expect that, in the next
decade, realistic models of cubic type non-Gaussianity will be tested
with real observations.

\section*{Acknowledgements}

We thank Paolo Creminelli and Leonardo Senatore for useful
discussions. We acknowledge support from the Swiss National Foundation
under contract No.  200021-116696/1 and WCU grant R32-2008-000-10130-0.

\appendix

\section{Non-Gaussian bias in the high peak limit}
\label{app:ngbias}

In this Appendix, we detail the derivation of the scale-dependent bias
correction induced by the $\gnl$ coupling to the two-point
correlation of dark matter haloes. We follow
\cite{2008ApJ...677L..77M} and  approximate the latter by the two-point
correlation  $\xi_{\rm hh}(\vr)$ of regions of the smoothed density
field $\delta_M$ with a peak height $\nu\gg 1$.

\subsection{case $\gnl\neq 0$ only}

In this case, only the four-point correlations of the density field
contribute at the first order. The leading-order correction to
the correlation of tresholded regions thus reads as
\begin{equation}
\Delta\xi_{\rm hh}=\frac{\nu^4}
{\sigma^4}\left[\frac{1}{3}\xi^{(4)}(\vx_1,\vx_1,\vx_1,\vx_2)
+\frac{1}{4}\xi^{(4)}(\vx_1,\vx_1,\vx_2,\vx_2)\right]\;.
\end{equation}
The four-point correlations $\xi^{(4)}$ are Fourier transform of the
trispectrum of the density field, $T_\delta(\vk_1,\vk_2,\vk_3,\vk_4)$,
with conservation of momentum enforced. Linearly extrapolating the 
density field to present epoch, the latter can be expressed as
\begin{equation}
T_\delta(\vk_1,\vk_2,\vk_3,\vk_4)=\left(\prod_{i=1}^4\am(k_i)\right)\, 
T_\Phi(\vk_1,\vk_2,\vk_3,\vk_4)\;,
\label{eq:tridens}
\end{equation}
where the expression for the trispectrum of primordial curvature perturbation
\begin{equation}
T_\Phi(\vk_1,\vk_2,\vk_3,\vk_4)=6 \gnl\bigl[P_\phi(k_1) P_\phi(k_2)
P_\phi(k_3) + \mbox{(cyclic)} \bigr]
\label{eq:tri1st}
\end{equation}
follows straightforwardly from the Fourier mode relation
\begin{equation}
\Phi(\vk)=\phi(\vk)+\gnl\!\int\!\!\frac{d^3 k_1}{(2\pi)^3}\!\int\!\!
\frac{d^3 k_2}{(2\pi)^3}\phi(\vk_1)\phi(\vk_2)\phi(\vk-\vk_1-\vk_2)\;.
\end{equation}
Density and curvature perturbations are related through the Poisson
equation, whose scale-dependence is reflected in the function
$\am(k)=\alpha_R(k,z=0)$ defined in Eq.~(\ref{eq:alpha}).  Combining
these relations gives
\begin{widetext}
\begin{align}
\xi^{(4)}(\vx_1,\vx_1,\vx_1,\vx_2) &=  6 \gnl\left(
\prod_{i=1}^{3}\int\!\!\frac{d^3 k_i}{(2\pi)^3}\,\am(k_i)
P_\phi(k_i)\right)
\left[1+3\,\frac{P_\phi(k_{123})}{P_\phi(k_3)}\right]\am(k_{123})\,e^{i\vk_{123}\cdot\vr}
\\ \xi^{(4)}(\vx_1,\vx_1,\vx_2,\vx_2) &=  6 \gnl\left(
\prod_{i=1}^{3}\int\!\!\frac{d^3 k_i}{(2\pi)^3}\,\am(k_i)
P_\phi(k_i)\right)
\left[1+2\frac{P_\phi(k_{123})}{P_\phi(k_1)}+\frac{P_\phi(k_{123})}{P_\phi(k_3)}\right]
\am(k_{123})\, e^{i\vk_{12}\cdot\vr}\;,
\end{align}
\end{widetext}
where we have defined $\vk_{ij\cdots l}=\vk_i+\vk_j+\,\cdots\, +\vk_l$
for shorthand  convenience.  Since we will examine the effect of
non-Gaussianity on Fourier space statistics only, we take the Fourier
transform of the four-point functions. After some  simplification, we
arrive at
\begin{widetext}
\begin{align}
\int\!\! d^3 r\,\xi^{(4)}(\vx_1,\vx_1,\vx_1,\vx_2)e^{-i\vk\cdot\vr}
  &= 6 \gnl \am(k) P_\phi(k)
  \int\!\!\frac{d^3 k_1}{(2\pi)^3}\am(k_1) P_\phi(k_1)
  \int\!\!\frac{d^3 k_2}{(2\pi)^3}\am(k_2) P_\phi(k_2)\nonumber \\ 
  & \qquad \times \am\bigl(|\vk+\vk_{12}|\bigr)
  \left[3+\frac{P_\phi\bigl(|\vk+\vk_{12}|\bigr)}{P_\phi(k)}\right] \\
\int\!\! d^3 r\,\xi^{(4)}(\vx_1,\vx_1,\vx_2,\vx_2)e^{-i\vk\cdot\vr}
  &= 6 \gnl\int\!\!\frac{d^3 k_1}{(2\pi)^3}\am(k_1)
  \am(|\vk+\vk_1|)P_\phi(k_1) P_\phi\bigl(|\vk+\vk_1|\bigr)
  \int\!\!\frac{d^3 k_2}{(2\pi)^3}\am(k_2)\nonumber \\
  & \qquad \times \am\bigl(|\vk+\vk_2|\bigr) P_\phi(k_2)
  \left[1+2\frac{P_\phi\bigl(|\vk+\vk_2|\bigr)}
    {P_\phi\bigl(|\vk+\vk_1|\bigr)}
    +\frac{P_\phi\bigl(|\vk+\vk_2|\bigr)}{P_\phi(k_2)}\right]\;.
\end{align}
\end{widetext}
For realistic values of the spectral index ($n_s\sim 1$), the products
$\am(|\vk+\vk_i|)P_\phi(|\vk+\vk_i|)$ appearing in the  right-hand
side of the above equalities  formally diverges whenever $\vk+\vk_1=0$
due to the ultraviolet divergence of $P_\phi(k)$. To cure this
problem, one can set $P_\phi(k)=0$ for sufficiently small wavenumbers
or excise a thin shell centred at wavenumber $k_i$ from the integral.
In the large-scale limit $k\ll k_i$, the ratio
$P_\phi(|\vk+\vk_i|)/P_\phi(k)$ vanishes whereas
$P_\phi(|\vk+\vk_i|)/P_\phi(k_i)$ tends towards unity. In this case,
the above  expressions reduce to
\begin{widetext}
\begin{multline}
\label{eq:x1112}
  \int\!\! d^3 r\,\xi^{(4)}(\vx_1,\vx_1,\vx_1,\vx_2)e^{-i\vk\cdot\vr} 
  \approx 3 \gnl\,\sigma^4 S_3^{(1)}\!(M)\,\am(k) P_\phi(k)  \\
  + 6\gnl\am(k)\int\!\!\frac{d^3 k_1}{(2\pi)^3}\am(k_1) 
  P_\phi(k_1)\int\!\!\frac{d^3 k_2}{(2\pi)^3}\am(k_2) P_\phi(k_2)
  \am(k_{12}) P_\phi(k_{12})\;,
\end{multline}
\end{widetext}
and
\begin{multline}
\label{eq:x1122}
\lefteqn{\int\!\! d^3 r\,\xi^{(4)}(\vx_1,\vx_1,\vx_2,\vx_2)
e^{-i\vk\cdot\vr}} \\
\approx 24 \gnl\,\sigma^2 \int\!\!\frac{d^3 k_1}{(2\pi)^3}
\,\am^2(k_1)P_\phi^2(k_1) \;.
\end{multline}
Only the first term in the right-hand side of Eq.(\ref{eq:x1112}) is
not well behaved in the limit $k\rightarrow 0$ where it becomes
proportional to $k^{n_s-2}$.  The second scales as $k^2$, while the
Fourier transform of $\xi^{(4)}(\vx_1,\vx_1,\vx_2,\vx_2)$ asymptotes to
a constant.  A similar decomposition also arises in $\fnl$ models. For
this type of non-Gaussianity,  the first order correction is furnished
by the three-point function $\xi^{(3)}(\vx_1,\vx_1,\vx_2)$, whose
Fourier transform can be split into the familiar term
$2\fnl\sigma^2\am(k)P_\phi(k)$, and a second piece given by
\begin{equation}
\frac{1}{2}\fnl \am(k) \int\!\!\frac{d^3 k_1}{(2\pi)^3}
\,\am^2(k_1)P_\phi^2(k_1)\;.
\end{equation}
In both quadratic and cubic local non-Gaussianity, the term
proportional to $\am(k)$ can be neglected since, at the pivot point
$k=k_0$, its magnitude relative to the term involving $\am(k)
P_\phi(k)$ is only ${\cal O}(0.01)$ and ${\cal O}(10^{-6})$,
respectively. Moreover, it decreases as one goes to larger scales. By
contrast, it is not so obvious how to handle the term
(\ref{eq:x1122}).  In the non-Gaussian halo power spectrum, this term
would appear  multiplied by $\nu^4/(4\sigma^4)$,
\begin{align}
\lefteqn{6\,\gnl b_{\rm L}^2\frac{\dc^2(z)}{\sigma^2}
\int\!\!\frac{d^3 k_1}{(2\pi)^3}\am^2(k_1)P_\phi^2(k_1)} \\
&\qquad\simeq 6\times 10^{-4} \gnl b_{\rm L}^2\,\frac{D^2(0)}{D^2(z)}
\left(\frac{M}{10^{13}\mdh}\right)^{0.375} \nonumber\;.
\end{align}
The approximation (second line) holds for $10^{13}\leq M\leq
10^{14}\mdh$.  For $\gnl=10^6$ and $b_{\rm L}\gtrsim 3$, this can be
much larger than the typical shot-noise correction applied to the halo
power spectra we measure in the simulations (see below). For
$\gnl=-10^6$, this will certainly produce a halo power spectrum which
is negative at sufficiently low wavenumber. It is plausible that
higher order counter-terms in the expansion Eq.(\ref{eq:corrlevel})
renormalises its value. However, such a calculation is beyond the
scope of this paper, so the simplest choice is to ignore this term in
the following of the analysis. Hence, we can approximate the
non-Gaussian correction $\Delta P_{\rm hh}$ in the limit of long
wavelength $k\ll 1$ by the Fourier transform of
$\nu^4\xi^{(4)}(\vx_1,\vx_1,\vx_2,\vx_2)/3\sigma^4$,
Eq.~(\ref{eq:dphhg}).

\subsection{case $\gnl$ and $\fnl$ non-zero}

In addition to the first order trispectrum induced by $\gnl\phi^3$,
Eq.~(\ref{eq:tri1st}), the quadratic coupling $\fnl\phi^2$ generates
the bispectrum at leading  order
\begin{equation}
B_\Phi(\vk_1,\vk_2,\vk_3) = 2\fnl\bigl[P(k_1)P(k_2)+\mbox{(cyclic)}
\bigr]\;,
\label{eq:bi1st}
\end{equation}
and an additional, albeit second order, contribution to the trispectrum
\cite{2006PhRvD..73b1301B,2006PhRvD..74l3519B,2006PhRvD..74l1301H},
\begin{align}
T_\Phi(\vk_1,\vk_2,\vk_3,\vk_4) &= 4\fnl^2\bigl[P_\phi(k_{13}) P_\phi(k_3)
P_\phi(k_4)\bigr. \nonumber \\ & 
\qquad \bigl. + 11 ~\mbox{permutations} \bigr] \;.
\label{eq:tri2nd}
\end{align}
The bispectrum (\ref{eq:bi1st}) induces a three-point contribution 
$(\nu^3/\sigma^3)\xi^{(3)}(\vx_1,\vx_1,\vx_2)$ to the power spectrum
of biased tracers which is calculated in \cite{2008ApJ...677L..77M}.
Upon Fourier transformation, it reads as
\begin{equation}
  \Delta\phh(k) = 4\fnl b_{\rm L}^2\dc(z)\am(k)P_\phi(k)\;.
\end{equation}
We follow the steps outlined above to calculate the contribution from
the second order trispectrum (\ref{eq:tri2nd}).  After some algebra,
the Fourier transform of the four-point correlations of the density
field can be written down as
\begin{widetext}
\begin{align}
\int\!\! d^3 r\,\xi^{(4)}(\vx_1,\vx_1,\vx_1,\vx_2)e^{-i\vk\cdot\vr}
  &= 8 \fnl^2 \am(k) P_\phi(k) \int\!\!\frac{d^3
  k_1}{(2\pi)^3}\am(k_1) P_\phi(k_1) \int\!\!\frac{d^3
  k_2}{(2\pi)^3}\am(k_2) P_\phi(k_2)\am\bigl(|\vk+\vk_{12}|\bigr)\nonumber \\
  & \quad\times\left\{\frac{P_\phi(k_{12})}{P_\phi(k_2)}
  \left[1+\frac{P_\phi\bigl(|\vk+\vk_{12}|\bigr)}{P_\phi(k)}\right]
  +\frac{P_\phi\bigl(|\vk+\vk_1|\bigr)}{P_\phi(k)}+
  \frac{P_\phi\bigl(|\vk+\vk_1|\bigr)}{P_\phi(k_1)} \right. \nonumber \\ 
  & \qquad\quad\left. +
  \frac{P_\phi\bigl(|\vk+\vk_1|\bigr)P_\phi\bigl(|\vk+\vk_{12}|\bigr)}
  {P_\phi(k_1)P_\phi(k_2)}+
  \frac{P_\phi\bigl(|\vk+\vk_1|\bigr)P_\phi\bigl(|\vk+\vk_{12}|\bigr)} 
       {P_\phi(k)P_\phi(k_2)}\right\}
\end{align}
and
\begin{align}
\int\!\! d^3 r\,\xi^{(4)}(\vx_1,\vx_1,\vx_2,\vx_2)e^{-i\vk\cdot\vr} &=
  4\fnl^2\int\!\!\frac{d^3 k_1}{(2\pi)^3}\am(k_1)
  \am\bigl(|\vk+\vk_1|\bigr)P_\phi(k_1)\int\!\!\frac{d^3 k_2}{(2\pi)^3}
  \am(k_2)\am\bigl(|\vk+\vk_2|\bigr)P_\phi(k_2) \nonumber \\ & 
  \quad\times \left\{P_\phi\bigl(|\vk+\vk_{12}|\bigr)
  \left[1+2\frac{P_\phi\bigl(|\vk+\vk_1|\bigr)}{P_\phi(k_2)}
    +\frac{P_\phi\bigl(|\vk+\vk_1|\bigr)P_\phi\bigl(|\vk+\vk_{2}|\bigr)}
    {P_\phi(k_1) P_\phi(k_2)}\right]\right.\nonumber \\ &
  \qquad\quad\left. +P_\phi(k)\left[1+2\frac{P_\phi\bigl(|\vk+\vk_1|\bigr)}
  {P_\phi(k_1)}+\frac{P_\phi\bigl(|\vk+\vk_1|\bigr)
    P_\phi\bigl(|\vk+\vk_{2}|\bigr)}
  {P_\phi(k_1) P_\phi(k_2)}\right] \right. \nonumber \\ &
  \qquad\quad\left. +2 P_\phi\bigl(|\vk+\vk_1|\bigr)\left[\frac{P_\phi(k_{12})}
  {P_\phi(k_1)}+\frac{P_\phi(k_{12})}{P_\phi(k_2)}\right]\right\}\;.
\end{align}
In the large-scale limit $k\rightarrow 0$, these expressions asymptote to
\begin{align}
  \int\!\! d^3 r\,\xi^{(4)}(\vx_1,\vx_1,\vx_1,\vx_2)e^{-i\vk\cdot\vr} 
  &\approx 4\fnl^2\,\sigma^4 S_3^{(1)}\!(M)\am(k) P_\phi(k)
  + 8\fnl^2\am(k)\int\!\!\frac{d^3 k_1}{(2\pi)^3}\am^2(k_1)P_\phi(k_1) 
  \nonumber \\
  & \qquad \times \int\!\!\frac{d^3 k_2}{(2\pi)^3}\am^2(k_2)\am(k_{12})
  \bigl[ P_\phi(k_1) P_\phi(k_2) + P_\phi(k_1) P_\phi(k_{12}) + 
    P_\phi^2(k_{12}) \bigr] \label{eq:fx1112} \\
  \int\!\! d^3 r\,\xi^{(4)}(\vx_1,\vx_1,\vx_2,\vx_2)e^{-i\vk\cdot\vr} 
  &\approx 16\fnl^2\sigma^4 P_\phi(k)+16\fnl^2\int\!\!\frac{d^3 k_1}
	   {(2\pi)^3}\,\am^2(k_1)\int\!\!\frac{d^3 k_2}{(2\pi)^3}\,
	   \am^2(k_2) P_\phi(k_2) P_\phi(k_{12})\nonumber \\
  & \qquad \times \bigl[ P_\phi(k_1) + P_\phi(k_2) \bigr]\;. 
\label{eq:fx1122}
\end{align}
\end{widetext}
Ignoring the second piece in the right-hand side of Eqs (\ref{eq:fx1112})
and (\ref{eq:fx1122}), the non-Gaussian correction to the halo power
spectrum is then easily recast as Eq.~(\ref{eq:dphhfg}).

\section{Signal-to-noise for the CMB trispectrum}
\label{app:noisetricmb}

The formalism for the CMB trispectrum has been established in
\cite{2001PhRvD..64h3005H,2002PhRvD..66f3008O}.  The invariance of the
4-point harmonic function of the CMB temperature anisotropy field
under the 4!  permutations of the coefficients $a_{l_i}^{m_i}$ imposes
constraints on the CMB trispectrum $T_{l_3 l_4}^{l_1l_2}(L)$ which can
be enforced by defining
\begin{align}
T_{l_3 l_4}^{l_1 l_2}(L) &= P_{l_3 l_4}^{l_1 l_2}(L)
+\left(2L+1\right) \sum_{L'}\biggl[ (-1)^{l_2+l_3} \biggr. \nonumber \\
& \qquad\biggl.\times\left\{\begin{array}{ccc} l_1 & l_2 & L \\ l_4 &
l_3 & L'\end{array}\right\} P_{l_1 l_3}^{l_2 l_4}(L') +(-1)^{L+L'}
\biggr. \nonumber \\ & \qquad\biggl.\times\left\{\begin{array}{ccc} l_1
& l_2 & L \\ l_3 & l_4 & L'\end{array}\right\} P_{l_1 l_4}^{l_3
l_2}(L') \biggr]\;,
\label{eq:Test}
\end{align}
where curly brackets are Wigner-6j symbols,
\begin{align}
P_{l_3 l_4}^{l_1 l_2}(L) &= t_{l_3 l_4}^{l_1 l_2}(L)
+ (-1)^{\sum_i l_i}\,t_{l_3 l_4}^{l_1 l_2}(L)+(-1)^{L+l_3+l_4} 
\nonumber \\
& \qquad \times t_{l_4 l_3}^{l_1 l_2}(L)
+ (-1)^{L+l_1+l_2}t_{l_3 l_4}^{l_2 l_1}(L)\;,
\label{eq:Pest}
\end{align}
and the reduced trispectrum $t_{l_3 l_4}^{l_1 l_2}(L)$ is  symmetric
under the exchange of its upper and lower indices and fully 
characterises the model.

The expansion coefficients $a_l^m$ are related to the primordial 
curvature perturbation $\Phi(\vk)$ through 
\begin{equation}
a_l^m=4\pi (-i)^l \int\!\!\frac{d^3 k}{(2\pi)^3}\Phi(\vk) g_{_{Tl}}(k)
Y_l^{m\star}(\kk)\;,
\label{eq:alm}
\end{equation}
where $g_{_{Tl}}(k)$ is the radiation transfer function. The reduced
trispectrum can be calculated from this relation once the four-point
function $T_\Phi(\vk_1,\vk_2,\vk_3,\vk_4)$ is specified. For a local
cubic coupling $\gnl\phi^3$,
\begin{align}
t_{l_3 l_4}^{l_1 l_2}(L) &= \int_0^\infty\!\! dr\,r^2 \beta_{l_2}(r)
\beta_{l_4}(r)
h_{_{l_1 L l_2}} h_{_{l_3 L l_4}} \nonumber \\
& \qquad \times \left[\mu_{l_1}(r)\beta_{l_3}(r)+\beta_{l_1}(r)\mu_{l_3}(r)
\right]
\end{align}
with
\begin{align}
\beta_l(r) &= \frac{2}{\pi}\int_0^\infty\!\! dk\,k^2 P_\phi(k) 
g_{_{Tl}}(k) j_l(kr) \\
\mu_l(r) &= \frac{2}{\pi}\int_0^\infty\!\! dk\, k^2 \gnl 
g_{_{Tl}}(k) j_l(kr)\;,
\end{align}
and
\begin{equation}
h_{_{l_1 L l_2}} =\frac{1}{\sqrt{4\pi}}\,\pi_{_{l_1 L l_2}}
\left(\begin{array}{ccc} l_1 & l_2 & L \\ 0 & 0 & 0\end{array}\right)\;.
\end{equation}
We also use the notation
\begin{equation}
\pi_{_{l_1 \cdots l_j}}=\sqrt{(2l_1+1)\times\cdots\times (2l_j+1)}\;.
\end{equation}
Note that most of the contribution to $t_{l_3 l_4}^{l_1 l_2}(L)$ comes
from a small volume centred at the comoving distance $r_\star$ to the
surface of last scattering.

The Sachs-Wolfe approximation $g_{_{Tl}}(k)\approx -j_l(k r_\star)/3$
valid at low multipoles $l\ll 100$ provides a useful
order-of-magnitude estimate
\cite{1994ApJ...430..447G,2001PhRvD..63f3002K,2002PhRvD..66f3008O,
2006PhRvD..73h3007K}. In this limit, we can approximate $\mu_l(r)$ as 
$-\gnl\delta_D(r-r_\star)/(3 r_\star^2)$ since we assume $\gnl$ 
independent of wavenumber. Hence, the reduced trispectrum simplifies to
\begin{equation}
t_{l_3 l_4}^{l_1 l_2}(L)\approx 9\gnl C_{l_2}^{\rm SW} C_{l_4}^{\rm SW}
\left(C_{l_1}^{\rm SW}+C_{l_3}^{\rm SW}\right)h_{_{l_1 L l_2}} 
h_{_{l_3 L l_4}}\;,
\label{eq:reducedsw}
\end{equation}
Inserting this expression successively into eqs (\ref{eq:Pest}) and
(\ref{eq:Test}), the CMB trispectrum eventually reads as
\begin{align}
\label{eq:trisw}
T_{l_3 l_4}^{l_1 l_2}(L) &= \frac{27}{2\pi}\gnl\left(2L+1\right)
\pi_{_{l_1 l_2 l_3 l_4}}^2 \\ & \quad \times \left(\begin{array}{ccc}
l_1 & l_2 & L \\ 0 & 0 & 0 \end{array}\right)\left(\begin{array}{ccc}
l_3 & l_4 & L \\ 0 & 0 & 0 \end{array}\right) \nonumber \\ & \quad
\times \Bigl[ C_{l_1}^{\rm SW} C_{l_2}^{\rm SW} C_{l_3}^{\rm SW} +
\mbox{(cyclic)} \Bigr] \nonumber \;.
\end{align}
where
\begin{equation}
C_l^{\rm SW}=\frac{2}{9\pi}\int_0^\infty\!\! dk\,k^2 P_\phi(k)
j_l^2(k r_\star)\approx \frac{2\pi A_\phi}{9 l (l+1)}\;.
\end{equation}
The last equality assumes a nearly scale-invariant spectrum
$n_s\approx 1$.  The following relation between the Wigner-3j and 6j
symbols (e.g., Appendix  A of \cite{2001PhRvD..64h3005H}),
\begin{multline}
\sum_{l_3'} (2l_3'+1)(-1)^{\Sigma+l_3'-l_3-m_1-m_1'}
\left\{\begin{array}{ccc} l_1 & l_2 & l_3 \\ l_1' & l_2' & l_3'
\end{array}\right\} \\
\times
\left(\begin{array}{ccc} l_2 & l_1' & l_3' \\ m_2 & m_1' & -m_3' 
\end{array}\right)
\left(\begin{array}{ccc} l_1 & l_3' & l_2' \\ m_1 & m_3 & -m_2' 
\end{array}\right) \\
= \left(\begin{array}{ccc} l_1 & l_2 & l_3 \\ m_1 & m_2 & -m 
\end{array}\right)
\left(\begin{array}{ccc} l_3 & l_1' & l_2' \\ m & m_1' & -m_2' 
\end{array}\right)
\end{multline}
where $\Sigma=l_1+l_2+l_1'+l_2'$ and the value of $m$ is set by the
triangle condition,  can be useful to derive Eq.(\ref{eq:trisw}). The
signal-to-noise summed up to multipole $l_{\rm max}$,
Eq.~(\ref{eq:sntri}), then becomes
\begin{align}
\lefteqn{\left(\frac{S}{N}\right)^2\!\!(<l_{\rm max})=\left(\frac{27}
{2\pi}\right)^2
\gnl^2\sum_{l_1>l_2>l_3>l_4}^{l_{\rm max}}\pi_{_{l_1 l_2 l_3 l_4}}^2} 
& \nonumber \label{eq:snlsum} \\
& \qquad \times
\frac{\bigl[C_{l_1}^{\rm SW} C_{l_2}^{\rm SW}C_{l_3}^{\rm SW}
+\mbox{(cyclic)}\bigr]^2}
{C_{l_1}^{\rm SW}C_{l_2}^{\rm SW}C_{l_3}^{\rm SW}C_{l_4}^{\rm SW}} \\
& \qquad \times
\sum_{L=0}^{2 l_{\rm max}}(2L+1)\left(\begin{array}{ccc} l_1 & l_2 & L \\
  0 & 0 & 0 \end{array}\right)^2\left(\begin{array}{ccc} l_3 & l_4 & L \\
  0 & 0 & 0 \end{array}\right)^2\nonumber\;.
\end{align}
We can recast the sum over the diagonal modes $L$ into a manifestly
symmetric form with the  aid of the Gaunt integral
\begin{equation}
\frac{1}{2}\int_{-1}^{+1}\!\! dx\, P_{l_1}(x) P_{l_2}(x) P_{l_3}(x) =
\left(\begin{array}{ccc} l_2 & l_2 & l_3 \\ 0 & 0 & 0 \end{array}\right)^2
\end{equation}
and the orthogonality relation
\begin{equation}
\sum_{k=0}^{\infty} (2k+1) P_k(x) P_k(y) = 2\delta_D(x-y)\;,
\end{equation}
where $\delta_D$ is the Dirac delta. We find
\begin{align}
\lefteqn{\sum_{L=0}^{2l_{\rm max}}(2L+1)
\left(\begin{array}{ccc} l_1 & l_2 & L \\
  0 & 0 & 0 \end{array}\right)^2\left(\begin{array}{ccc} l_3 & l_4 & L \\
  0 & 0 & 0 \end{array}\right)^2} \\
&\qquad =\frac{1}{2}\int_{-1}^{+1}\!\! dx\, P_{l_1}(x) P_{l_2}(x) P_{l_3}(x) 
P_{l_4}(x) \nonumber \;.
\end{align}
There is a strict equality because the Wigner-3j symbols vanish for
$L>l_1+l_2$. Eq.~(\ref{eq:snsw}) for the signal-to-noise then follows
by replacing the above equality into Eq.(\ref{eq:snlsum}) and summing
over all the 4!  permutations of the quadruplet
$(l_1,l_2,l_3,l_4)$. Although Eq.~(\ref{eq:snsw}) becomes
computationally expensive when $l_{\rm max}\gg 100$ (because we are
summing over redundant configurations), we found that it is quite
efficient for $l_{\rm max}\lesssim 200$.

Following \cite{2006PhRvD..73h3007K}, we can roughly estimate the
dependence of the signal-to-noise squared on $l_{\rm max}$ by
considering only the contribution of the $L=1$ mode in
Eq.(\ref{eq:snlsum}). Consequently, the product of the Wigner-3j
symbols squared reduces to $\sim l_1 l_3\,
\delta_{l_1-1,l_2}\delta_{l_3-1,l_4}$ and yields $(S/N)^2\propto
l_{\rm max}^2$. However, including all $L$ modes as in
Eq.~(\ref{eq:snsw}) gives a steeper dependence, $(S/N)^2\propto l_{\rm
max}^{2.6}$ (see Fig.\ref{fig:sncmb}), due to the fact that the
Wigner-3j symbols decay slowly with increasing $L$. This is  quite
apparent in the classical limit $l_1,l_2,L\gg 1$, where
\begin{align}
\lefteqn{\left(\begin{array}{ccc} l_1 & l_2 & L \\
  0 & 0 & 0 \end{array}\right)^2} && \\
& \qquad \approx \frac{1}{\pi}
\bigl[(l_1+l_2)^2-L^2\bigr]^{-1/2}\bigl[L^2-(l_1-l_2)^2\bigr]^{-1/2}
\nonumber\;.
\end{align}
Clearly, the terms in the summation Eq.(\ref{eq:snlsum}) decay only as
$1/L$ for $L\gg l_1,l_2$. By contrast, the second  order contribution
to the CMB trispectrum induced by the quadratic coupling $\fnl\phi^2$
adds an additional multiplicative factor  of $(C_L^{\rm SW})^2$ in the
summation over the $L$ modes which increases the relative contribution
of the low-$L$ modes (since these now  decay as $L^{-5}$). This is the
reason why considering only $L\leq 10$ modes as done in
\cite{2006PhRvD..73h3007K} still provides a good approximation to the
signal-to-noise of the CMB trispectrum  for the $\fnl$ model.

\bibliographystyle{prsty}
\bibliography{gnl}

\begin{thebibliography}{10}

\bibitem{1981JETPL..33..532M}
V.~F. {Mukhanov} and G.~V. {Chibisov}, Soviet Journal of Experimental and
  Theoretical Physics Letters {\bf 33},  532  (1981).

\bibitem{1982PhLB..117..175S}
A.~A. {Starobinsky}, Physics Letters B {\bf 117},  175  (1982).

\bibitem{1982PhLB..115..295H}
S.~W. {Hawking}, Physics Letters B {\bf 115},  295  (1982).

\bibitem{1982PhRvL..49.1110G}
A.~H. {Guth} and S. {Pi}, Physical Review Letters {\bf 49},  1110  (1982).

\bibitem{1987PhLB..197...66A}
T.~J. {Allen}, B. {Grinstein}, and M.~B. {Wise}, Physics Letters B {\bf 197},
  66  (1987).

\bibitem{1992PhRvD..46.4232F}
T. {Falk}, R. {Rangarajan}, and M. {Srednicki}, \prd {\bf 46},  4232  (1992).

\bibitem{1994ApJ...430..447G}
A. {Gangui}, F. {Lucchin}, S. {Matarrese}, and S. {Mollerach}, \apj {\bf 430},
  447  (1994).

\bibitem{2004PhR...402..103B}
N. {Bartolo}, E. {Komatsu}, S. {Matarrese}, and A. {Riotto}, Phys. Rep. {\bf
  402},  103  (2004).

\bibitem{2003NuPhB.667..119A}
V. {Acquaviva}, N. {Bartolo}, S. {Matarrese}, and A. {Riotto}, Nuclear Physics
  B {\bf 667},  119  (2003).

\bibitem{2003JHEP...05..013M}
J. {Maldacena}, Journal of High Energy Physics {\bf 5},  13  (2003).

\bibitem{2003ApJS..148..119K}
E. {Komatsu}, A. {Kogut}, M.~R. {Nolta}, C.~L. {Bennett}, M. {Halpern}, G.
  {Hinshaw}, N. {Jarosik}, M. {Limon}, S.~S. {Meyer}, L. {Page}, D.~N.
  {Spergel}, G.~S. {Tucker}, L. {Verde}, E. {Wollack}, and E.~L. {Wright},
  "Astrophys. J. Supp." {\bf 148},  119  (2003).

\bibitem{2007JCAP...03..005C}
P. {Creminelli}, L. {Senatore}, M. {Zaldarriaga}, and M. {Tegmark}, Journal of
  Cosmology and Astro-Particle Physics {\bf 3},  5  (2007).

\bibitem{2009ApJS..180..330K}
E. {Komatsu}, J. {Dunkley}, M.~R. {Nolta}, C.~L. {Bennett}, B. {Gold}, G.
  {Hinshaw}, N. {Jarosik}, D. {Larson}, M. {Limon}, L. {Page}, D.~N. {Spergel},
  M. {Halpern}, R.~S. {Hill}, A. {Kogut}, S.~S. {Meyer}, G.~S. {Tucker}, J.~L.
  {Weiland}, E. {Wollack}, and E.~L. {Wright}, ApJS {\bf 180},  330  (2009).

\bibitem{2009arXiv0901.2572S}
K.~M. {Smith}, L. {Senatore}, and M. {Zaldarriaga}, ArXiv e-prints  (2009).

\bibitem{2009MNRAS.393..615C}
A. {Curto}, E. {Mart{\'{\i}}nez-Gonz{\'a}lez}, P. {Mukherjee}, R.~B.
  {Barreiro}, F.~K. {Hansen}, M. {Liguori}, and S. {Matarrese}, Mon. Not. R.
  Astron. Soc. {\bf 393},  615  (2009).

\bibitem{2003PhRvD..67b3503L}
D.~H. {Lyth}, C. {Ungarelli}, and D. {Wands}, \prd {\bf 67},  023503  (2003).

\bibitem{2004PhRvD..69d3503B}
N. {Bartolo}, S. {Matarrese}, and A. {Riotto}, \prd {\bf 69},  043503  (2004).

\bibitem{2005JCAP...10..013E}
K. {Enqvist} and S. {Nurmi}, Journal of Cosmology and Astro-Particle Physics
  {\bf 10},  13  (2005).

\bibitem{2006JCAP...09..008M}
K.~A. {Malik} and D.~H. {Lyth}, Journal of Cosmology and Astro-Particle Physics
  {\bf 9},  8  (2006).

\bibitem{2006PhRvD..74j3003S}
M. {Sasaki}, J. {V{\"a}liviita}, and D. {Wands}, \prd {\bf 74},  103003
  (2006).

\bibitem{2008JCAP...09..012E}
K. {Enqvist} and T. {Takahashi}, Journal of Cosmology and Astro-Particle
  Physics {\bf 9},  12  (2008).

\bibitem{2008JCAP...09..025H}
Q.-G. {Huang} and Y. {Wang}, Journal of Cosmology and Astro-Particle Physics
  {\bf 9},  25  (2008).

\bibitem{2008PhRvD..78b3513I}
K. {Ichikawa}, T. {Suyama}, T. {Takahashi}, and M. {Yamaguchi}, \prd {\bf 78},
  023513  (2008).

\bibitem{2009JCAP...04..031C}
P. {Chingangbam} and Q.-G. {Huang}, Journal of Cosmology and Astro-Particle
  Physics {\bf 4},  31  (2009).

\bibitem{2008JCAP...11..005H}
Q.-G. {Huang}, Journal of Cosmology and Astro-Particle Physics {\bf 11},  5
  (2008).

\bibitem{2009JCAP...06..035H}
Q.-G. {Huang}, Journal of Cosmology and Astro-Particle Physics {\bf 6},  35
  (2009).

\bibitem{2009JCAP...08..016B}
C.~T. {Byrnes} and G. {Tasinato}, Journal of Cosmology and Astro-Particle
  Physics {\bf 8},  16  (2009).

\bibitem{2009PhRvD..80f3503L}
J.-L. {Lehners} and S. {Renaux-Petel}, \prd {\bf 80},  063503  (2009).

\bibitem{2001PhRvD..64h3005H}
W. {Hu}, \prd {\bf 64},  083005  (2001).

\bibitem{2002PhRvD..66f3008O}
T. {Okamoto} and W. {Hu}, Phys. Rev. D {\bf 66},  063008  (2002).

\bibitem{2006PhRvD..73h3007K}
N. {Kogo} and E. {Komatsu}, Phys. Rev. D {\bf 73},  083007  (2006).

\bibitem{2002astro.ph..6039K}
E. {Komatsu}, ArXiv Astrophysics e-prints  (2002).

\bibitem{2001ApJ...563L..99K}
M. {Kunz}, A.~J. {Banday}, P.~G. {Castro}, P.~G. {Ferreira}, and K.~M.
  {G{\'o}rski}, Astrophys. J. Lett. {\bf 563},  L99  (2001).

\bibitem{1988ApJ...330..535L}
F. {Lucchin} and S. {Matarrese}, \apj {\bf 330},  535  (1988).

\bibitem{1989ApJ...345....3C}
S. {Colafrancesco}, F. {Lucchin}, and S. {Matarrese}, \apj {\bf 345},  3
  (1989).

\bibitem{1986ApJ...310...19G}
B. {Grinstein} and M.~B. {Wise}, \apj {\bf 310},  19  (1986).

\bibitem{1986ApJ...310L..21M}
S. {Matarrese}, F. {Lucchin}, and S.~A. {Bonometto}, Astrophys. J. Lett. {\bf
  310},  L21  (1986).

\bibitem{2004PhRvD..69j3513S}
R. {Scoccimarro}, E. {Sefusatti}, and M. {Zaldarriaga}, \prd {\bf 69},  103513
  (2004).

\bibitem{2007PhRvD..76h3004S}
E. {Sefusatti} and E. {Komatsu}, \prd {\bf 76},  083004  (2007).

\bibitem{2009arXiv0905.0717S}
E. {Sefusatti}, ArXiv e-prints  (2009).

\bibitem{2009arXiv0904.0497J}
D. {Jeong} and E. {Komatsu}, ArXiv e-prints  (2009).

\bibitem{2009JCAP...01..010K}
M. {Kamionkowski}, L. {Verde}, and R. {Jimenez}, Journal of Cosmology and
  Astro-Particle Physics {\bf 1},  10  (2009).

\bibitem{2009MNRAS.395.1743L}
T.~Y. {Lam} and R.~K. {Sheth}, Mon. Not. R. Astron. Soc. {\bf 395},  1743
  (2009).

\bibitem{2006ApJ...653...11H}
C. {Hikage}, E. {Komatsu}, and T. {Matsubara}, \apj {\bf 653},  11  (2006).

\bibitem{2008MNRAS.385.1613H}
C. {Hikage}, P. {Coles}, M. {Grossi}, L. {Moscardini}, K. {Dolag}, E.
  {Branchini}, and S. {Matarrese}, Mon. Not. R. Astron. Soc. {\bf 385},  1613
  (2008).

\bibitem{2008PhRvD..77l3514D}
N. {Dalal}, O. {Dor{\'e}}, D. {Huterer}, and A. {Shirokov}, \prd {\bf 77},
  123514  (2008).

\bibitem{2008ApJ...677L..77M}
S. {Matarrese} and L. {Verde}, Astrophys. J. Lett. {\bf 677},  L77  (2008).

\bibitem{2008JCAP...08..031S}
A. {Slosar}, C. {Hirata}, U. {Seljak}, S. {Ho}, and N. {Padmanabhan}, Journal
  of Cosmology and Astro-Particle Physics {\bf 8},  31  (2008).

\bibitem{2008PhRvD..78l3507A}
N. {Afshordi} and A.~J. {Tolley}, \prd {\bf 78},  123507  (2008).

\bibitem{2008ApJ...684L...1C}
C. {Carbone}, L. {Verde}, and S. {Matarrese}, Astrophys. J. Lett. {\bf 684},
  L1  (2008).

\bibitem{2009PhRvL.102b1302S}
U. {Seljak}, Physical Review Letters {\bf 102},  021302  (2009).

\bibitem{2009arXiv0906.0232S}
E. {Sefusatti}, M. {Liguori}, A.~P.~S. {Yadav}, M.~G. {Jackson}, and E.
  {Pajer}, ArXiv e-prints  (2009).

\bibitem{2009MNRAS.396...85D}
V. {Desjacques}, U. {Seljak}, and I.~T. {Iliev}, Mon. Not. R. Astron. Soc. {\bf
  396},  85  (2009).

\bibitem{2008arXiv0811.4176P}
A. {Pillepich}, C. {Porciani}, and O. {Hahn}, ArXiv e-prints  (2008).

\bibitem{2009arXiv0902.2013G}
M. {Grossi}, L. {Verde}, C. {Carbone}, K. {Dolag}, E. {Branchini}, F.
  {Iannuzzi}, S. {Matarrese}, and L. {Moscardini}, ArXiv e-prints  (2009).

\bibitem{1996ApJ...469..437S}
U. {Seljak} and M. {Zaldarriaga}, \apj {\bf 469},  437  (1996).

\bibitem{2005MNRAS.364.1105S}
V. {Springel}, Mon. Not. R. Astron. Soc. {\bf 364},  1105  (2005).

\bibitem{1970A&A.....5...84Z}
Y.~B. {Zel'Dovich}, A\&A {\bf 5},  84  (1970).

\bibitem{1984ApJ...279..499F}
J.~N. {Fry}, \apj {\bf 279},  499  (1984).

\bibitem{1986ApJ...311....6G}
M.~H. {Goroff}, B. {Grinstein}, S.-J. {Rey}, and M.~B. {Wise}, \apj {\bf 311},
  6  (1986).

\bibitem{1992ApJ...394L...5B}
F.~R. {Bouchet}, R. {Juszkiewicz}, S. {Colombi}, and R. {Pellat}, Astrophys. J.
  Lett. {\bf 394},  L5  (1992).

\bibitem{1994ApJ...433....1B}
F. {Bernardeau}, \apj {\bf 433},  1  (1994).

\bibitem{1998MNRAS.299.1097S}
R. {Scoccimarro}, Mon. Not. R. Astron. Soc. {\bf 299},  1097  (1998).

\bibitem{2008PhRvD..78l3519M}
P. {McDonald}, \prd {\bf 78},  123519  (2008).

\bibitem{2009arXiv0904.3662K}
S.~R. {Knollmann} and A. {Knebe}, ArXiv e-prints  (2009).

\bibitem{2004MNRAS.351..399G}
S.~P.~D. {Gill}, A. {Knebe}, and B.~K. {Gibson}, Mon. Not. R. Astron. Soc. {\bf
  351},  399  (2004).

\bibitem{1992ApJ...399..405W}
M.~S. {Warren}, P.~J. {Quinn}, J.~K. {Salmon}, and W.~H. {Zurek}, \apj {\bf
  399},  405  (1992).

\bibitem{1994MNRAS.271..676L}
C. {Lacey} and S. {Cole}, Mon. Not. R. Astron. Soc. {\bf 271},  676  (1994).

\bibitem{1996MNRAS.282..263E}
V.~R. {Eke}, S. {Cole}, and C.~S. {Frenk}, Mon. Not. R. Astron. Soc. {\bf 282},
   263  (1996).

\bibitem{1998ApJ...495...80B}
G.~L. {Bryan} and M.~L. {Norman}, \apj {\bf 495},  80  (1998).

\bibitem{1974ApJ...187..425P}
W.~H. {Press} and P. {Schechter}, \apj {\bf 187},  425  (1974).

\bibitem{1991ApJ...379..440B}
J.~R. {Bond}, S. {Cole}, G. {Efstathiou}, and N. {Kaiser}, \apj {\bf 379},  440
   (1991).

\bibitem{2008JCAP...04..014L}
M. {Lo Verde}, A. {Miller}, S. {Shandera}, and L. {Verde}, Journal of Cosmology
  and Astro-Particle Physics {\bf 4},  14  (2008).

\bibitem{2000ApJ...541...10M}
S. {Matarrese}, L. {Verde}, and R. {Jimenez}, \apj {\bf 541},  10  (2000).

\bibitem{2007MNRAS.382.1261G}
M. {Grossi}, K. {Dolag}, E. {Branchini}, S. {Matarrese}, and L. {Moscardini},
  Mon. Not. R. Astron. Soc. {\bf 382},  1261  (2007).

\bibitem{1999MNRAS.308..119S}
R.~K. {Sheth} and G. {Tormen}, Mon. Not. R. Astron. Soc. {\bf 308},  119
  (1999).

\bibitem{2009arXiv0906.1042V}
P. {Valageas}, ArXiv e-prints  (2009).

\bibitem{2009arXiv0905.1702L}
T.~Y. {Lam} and R.~K. {Sheth}, ArXiv e-prints  (2009).

\bibitem{2009MNRAS.399.1482L}
T.~Y. {Lam}, R.~K. {Sheth}, and V. {Desjacques}, Mon. Not. R. Astron. Soc. {\bf
  399},  1482  (2009).

\bibitem{2009JCAP...03..004S}
A. {Slosar}, Journal of Cosmology and Astro-Particle Physics {\bf 3},  4
  (2009).

\bibitem{2009PhRvL.103i1303S}
U. {Seljak}, N. {Hamaus}, and V. {Desjacques}, Physical Review Letters {\bf
  103},  091303  (2009).

\bibitem{2008arXiv0810.0323M}
P. {McDonald} and U. {Seljak}, ArXiv e-prints  (2008).

\bibitem{2001PhRvD..63f3002K}
E. {Komatsu} and D.~N. {Spergel}, \prd {\bf 63},  063002  (2001).

\bibitem{2008MNRAS.390..438G}
M. {Grossi}, E. {Branchini}, K. {Dolag}, S. {Matarrese}, and L. {Moscardini},
  Mon. Not. R. Astron. Soc. {\bf 390},  438  (2008).

\bibitem{2008PhRvD..78l3534T}
A. {Taruya}, K. {Koyama}, and T. {Matsubara}, \prd {\bf 78},  123534  (2008).

\bibitem{2007MNRAS.376..343K}
X. {Kang}, P. {Norberg}, and J. {Silk}, Mon. Not. R. Astron. Soc. {\bf 376},
  343  (2007).

\bibitem{2002MNRAS.329...61S}
R.~K. {Sheth} and G. {Tormen}, Mon. Not. R. Astron. Soc. {\bf 329},  61
  (2002).

\bibitem{2009arXiv0903.1251M}
M. {Maggiore} and A. {Riotto}, ArXiv e-prints  (2009).

\bibitem{2009ApJ...692..217L}
Z. {Luki{\'c}}, D. {Reed}, S. {Habib}, and K. {Heitmann}, \apj {\bf 692},  217
  (2009).

\bibitem{2006PhRvD..73b1301B}
L. {Boubekeur} and D.~H. {Lyth}, \prd {\bf 73},  021301  (2006).

\bibitem{2006PhRvD..74l3519B}
C.~T. {Byrnes}, M. {Sasaki}, and D. {Wands}, \prd {\bf 74},  123519  (2006).

\bibitem{2006PhRvD..74l1301H}
M.-X. {Huang} and G. {Shiu}, \prd {\bf 74},  121301  (2006).

\end{thebibliography}

\end{document}